# Phonon-Enabled Collective Strong Coupling of Color Centers in a Plasmonic Cavity

Bharti Garg†, Prabhdeep Singh†, Masoud Taleb, Nahid Talebi*

Institute for Experimental and Applied Physics, Kiel University, Germany

Email: talebi@physik.uni-kiel.de

† These authors contributed equally to this work.

**Abstract**: Quantum emitters in hexagonal boron nitride (hBN) are promising building blocks for solid-state quantum photonics because of their bright room-temperature emission. These emitters strongly interact with lattice vibrations, which partially underpins their excitation efficiency. Here, using complementary cathodoluminescence (CL) and photoluminescence (PL) spectroscopy, we demonstrate cavity-mediated interactions between hBN quantum emitters coupled to the same plasmonic Au void cavity at the room temperature. Cathodoluminescence measurements reveal that a vibronic transition of the 443 nm emitter hybridizes with the cavity mode, giving rise to a spectrally resolved splitting, while the neighboring 537 nm emitter remains in the weak-coupling regime despite spectral overlap. The energy split of the CL peak is at the order of 36.2 meV, whereas the PL-resolved peak is approximately 24.1 meV, which is due to the collective excitations of multiple emitters per electron excitation and their mutual interactions. Maxwell-Bloch simulations reproduce the experimental spectra and identify the observed splitting as the result of cavity-induced hybridization. Excitation-dependent PL further shows that the hybridized state can be accessed through phonon-assisted excitations below the zero-phonon transition. These results demonstrate that vibronic transitions provide an effective pathway for controlling interactions between multiple quantum emitters in hybrid plasmonic nanophotonic systems.

## Introduction

Semiconductor quantum dots (QDs) and diamond color centers have long served as benchmark solid-state platforms for single-photon generation and cavity quantum electrodynamics as they enable bright non-classical light emission and strong light-matter interactions[1-6]. One of the major challenges in cavity quantum electrodynamics is achieving the strong-coupling regime, in which the emitter-cavity coupling strength exceeds the combined dissipative losses of the system, allowing coherent energy exchange between the emitter and the cavity[7-10]. Accessing this regime requires optical cavities that simultaneously provide confined optical mode volumes and low optical losses. Photonic cavities such as micropillars provide such a confined model volume. In addition, simultaneous excitations of multiple emitters that lead to a larger collective coupling strength and hence a larger spectrally-resolved energy split has been reported in the realm of semiconductor QDs and superconducting qubits. However, semiconductor QDs[11-13] and superconducting qubits[14] often require cryogenic operation to preserve optical coherence and are difficult to integrate deterministically into on-chip photonic architectures[15-18].

In recent years, hexagonal boron nitride (hBN) has emerged as a versatile two-dimensional van der Waals material owing to its layered crystal structure and exceptional optical properties, supporting bright and stable single-photon emitters[19,20] (SPEs) that operate from cryogenic to above room temperatures[21,22]. These SPEs, commonly associated with lattice defects in hBN, exhibit a vibronic structure characterized by a narrow zero-phonon line (ZPL) and significant phonon sidebands (PSBs)[23,24]. The coupling of electronic transitions to lattice vibrations is a defining feature of hBN photophysics[25]. Recent theoretical and experimental studies have shown that coherent phonon-assisted vibronic transitions can assist in exciting the zero-phonon transition, thereby enabling coherent optical control beyond direct resonant excitation[26,27]. Rather than serving solely as a dissipative environment, phonons can facilitate cavity-mediated interactions by enabling optical transitions that relax spectral matching constraints between emitters and resonant modes[28].

The photophysics of hBN SPEs and their coupling to phonons have been explored using both PL and CL[29-32]. Whereas laser excitations generally excite the system coherently and can further lead to additional features such as stimulated Raman scattering[27], CL unambiguously reveals the optical signatures of SPEs[32,33].

Despite these advances, current literature mostly targets the way individual ZPLs interact with specific cavity modes, usually in systems containing a single SPE[22,34-36]. While conventional approaches have heavily relied upon dielectric cavities, there is an increasing interest in hybrid geometries that couple hBN emitters to plasmonic nanostructures. For instance, integrating hBN with gold or silver nanoparticles in nanoparticle-on-mirror geometries or periodic arrays provides extreme field confinement and Purcell enhancement, significantly boosting emission intensity and modifying radiative lifetimes[37,38]. A significant gap remains in understanding how distinct coupling regimes coexist within a single cavity, particularly when a hybridized vibronic mode of one emitter spectrally overlaps with the photonic cavity modes. While phonon-mediated interactions have been theoretically investigated in quantum dots, analogous interactions involving hybridized vibronic modes and neighboring quantum emitters remain largely unexplored in hBN.

Here, using correlative CL and PL spectroscopy, we show that a vibronic transition of one hBN emitter hybridizes with a plasmonic cavity mode, while a neighboring emitter remains weakly coupled despite the spectral overlap of its ZPL with the cavity. Our emitter network constitutes two types of emitters: a "blue" center (Emitter 1) with a ZPL at $\lambda$ = 443 nm (2.8 eV) and a structural-defect-related center (Emitter 2) at $\lambda$ = 537 nm (2.31 eV). We identify Emitter 1 as a substituted vertical carbon dimer (VACD)[39] and Emitter 2 as an emission center related to arrays of dislocations[40]. We report the observation of strong vibronic-mode coupling, in which the 4$^{th}$-order vibronic mode of Emitter 1 hybridizes with the cavity mode to produce a resolved energy splitting centered around 550 nm in CL and 544 nm in PL. The observed CL-resolved energy split is at the order of 36.24 meV, whereas the PL-resolved split is measured to be approximately 24.09 meV. This difference is associated with the fact that electrons excite more emitters simultaneously, which leads to the collective strong-coupling of the vibronic modes of various emitters with the same cavity mode. Crucially, we show using Maxwell-Bloch simulations frameworks constituting a network of randomly-oriented emitters coupled to our plasmonic cavity, that Emitter 2 simultaneously resides in the weak-coupling regime, exhibiting no splitting despite a significant spectral overlap between its ZPL and the hybridized vibronic mode of Emitter 1. Furthermore, we demonstrate that the coherent down-converted laser beam caused by the stimulated Raman scattering can coherently interact with emitters, leading to a more efficient and coherent excitation scheme. Our results provide a unique platform for studying inter-emitter interactions mediated by a common plasmonic void cavity and highlights the potential for multi-SPEs quantum state manipulation in hybrid two-dimensional nanophotonic architectures.

## Results and Discussions

### Defect Natures and Optical Characteristics

The optical properties of defect centers in hBN are investigated here using complementary CL and PL spectroscopy. The measurements are performed on liquid-exfoliated hBN flakes with thicknesses ranging from a few nanometers to several tens of micrometers, deposited onto two types of substrates: (i) A free-standing Au plasmonic crystal composed of a square array of pin holes (commercially-available transmission electron microscopy grids), and (ii) a thermally deposited thin film of gold on a glass substrate, where both Au structures exhibit 40 nm thickness. Liquid exfoliation has previously been shown to produce stable quantum emitters in hBN[27,32,41] (see Methods for sample preparation). The experimental configuration is shown in Fig. 1a, illustrating the electron- and photon-excitation schemes used to investigate the hBN flakes on Au grid with a hole diameter of 600-800nm. The flake thickness and morphology are first characterized using a MarSurf CM Explorer confocal microscope and scanning electron microscopy (SEM), followed by CL and PL spectroscopy.

The CL spectra shown in Fig. 1c and d illustrate the characteristic optical response of the two main defect centers investigated in this work. The spectra reveal two distinct classes of visible quantum emitters, with their ZPL centered at approximately 443 nm and 537 nm. The 443 nm (2.80 eV) emitter is commonly designated as the B center[39,42,43]. Recent studies have identified its microscopic origin as a VACD occupying a split-interstitial configuration. The formation of the blue B center has been consistently observed under electron-beam irradiation, which has been shown to induce a one-to-one conversion of ultraviolet-emitting in-plane carbon dimers ($C_BC_N$) into blue B centers[44].

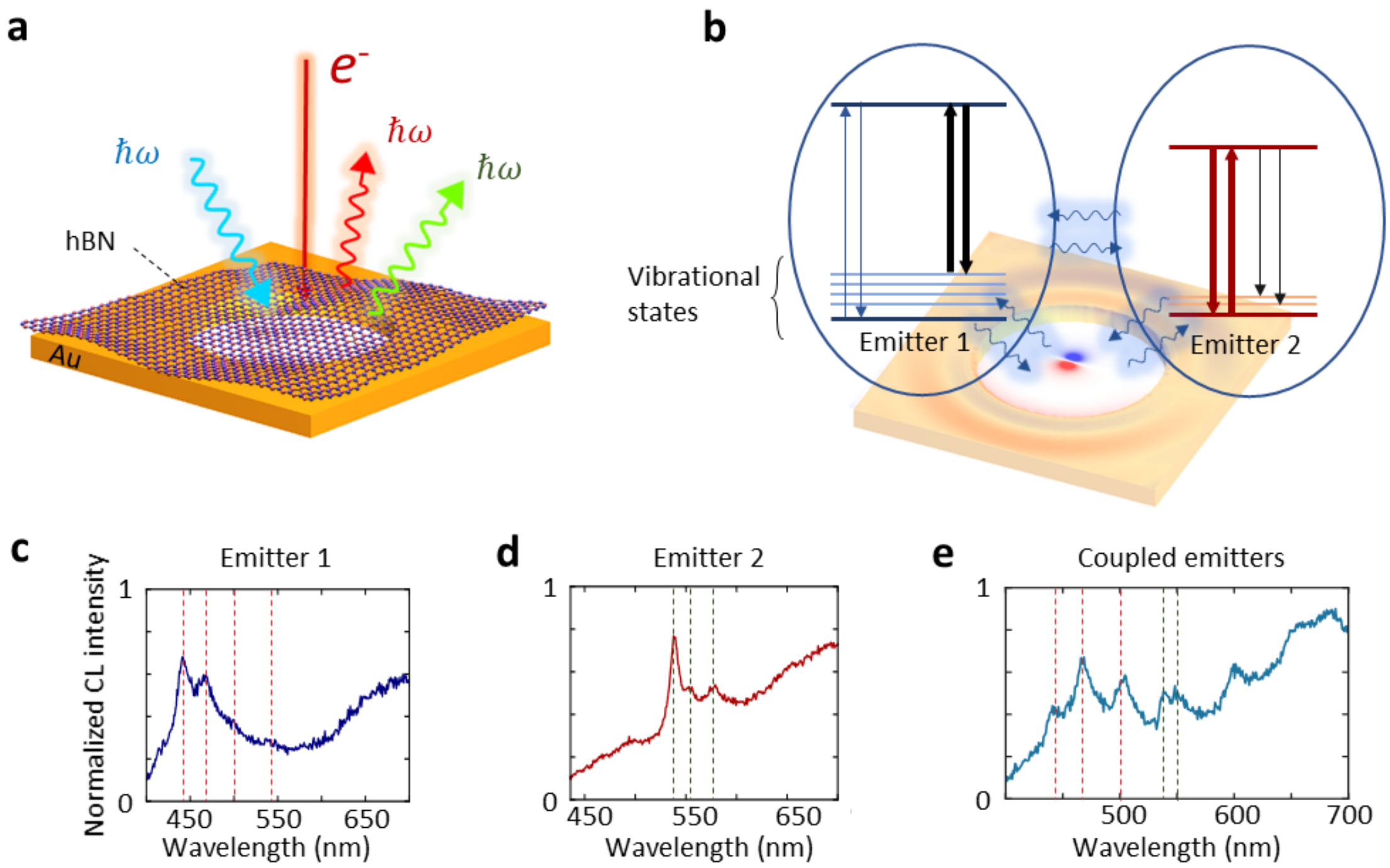

**Figure 1. Experimental concept and optical response of individual and cavity-coupled quantum emitters in hBN. (a)** Schematic of photon and electron excitation of hBN transferred onto an Au plasmonic void cavity, where the focused beams excite the hBN defect centers. **(b)** Conceptual energy-level diagram of the two emitters coupled through the confined optical mode of the plasmonic cavity. **(c** and **d)** Normalized CL spectra of the individual emitters centered at 443 nm (Emitter 1) and 537 nm (Emitter 2), with dashed lines indicating the ZPLs and their phonon sidebands. **(e)** CL spectrum of the cavity-coupled system, showing an energy split of the 4$^{th}$ vibrational resonance of Emitter 1.

The second emitter centered around 537 nm (2.31 eV) is generally associated with extended structural defects rather than isolated point defects. Although its microscopic origin is still under debate, several studies have attributed this emission to dislocation arrays and the strain generated during crystal deformation, where local strain creates deformation potential wells that trap charge carriers and activate pre-existing defect states[40,45]. Carbon-rich hBN has also been reported to host radiative centers emitting near 2.3 eV, suggesting that carbon-containing defect complexes may contribute to emissions around these wavelengths[46]. Since no intentional carbon incorporation was introduced during sample preparation, any carbon-related defects present in our samples are likely to originate from residual contamination introduced during the liquid exfoliation or synthesis itself. Additionally, argon plasma treatment was performed to reduce the background emission and enhance the luminescence intensity of the defect centers[47].

In addition to determining the emitter ZPL energies, the spectral linewidths of the emitters were carefully analyzed by fitting the emission lines with pseudo-Voigt functions[27,48]. FWHM linewidths for emitter 1 accounts for 14.4 nm, whereas emitter type 2 exhibited a narrower linewidth of approximately 11.8 nm.

**Cavity-Mediated Cathodoluminescence**

We next investigate the optical response of these emitters within a common cavity environment using CL. The electron kinetic energy is 20 keV. For more details about the CL method and parameters, see the Methods Section. The conceptual configuration is shown in Fig. 1b, where both emitters are positioned inside the highly confined electromagnetic field of an Au plasmonic void cavity. Owing to its ultrasmall mode volume, the cavity enhances the local density of optical states and provides a common optical mode to which both emitters can couple.

Supplementary Note 1 and supplementary Note 7 provide more information about the modal configurations of the void plasmonic structures. Particularly the so-called photonic modes of the structure, i.e., those confined within the hole rather than surface plasmon resonances, sustain a larger quality factor. Localized plasmonic excitations at the rim of the structure exhibit in contrast a larger localization and are responsible for the strong interactions we observe here[49]. Due to the large periodicity of the plasmonic lattice, collective responses of the periodic grating happen at extensively larger wavelengths compared to the emission wavelength of the emitters; hence, the emitters couple to only individual void-plasmon resonances.

The explored structure is therefore, composed of a thin hBN flake (75 nm; see the thickness map in Supplementary Note 8 and Supplementary Fig. S15) positioned on top of an Au grid. The Au grid exhibits a square array of pin holes, where the hole diameter is ranging between 600-800 nm and the lattice periodicity is 1600 nm (see Fig. 1a for SEM and optical images of the structure). Hyperspectral CL images unravel a high density of emitters positioned in the hBN flakes (see Fig. 2d, as well as Supplementary Fig. S8 to S11), a fact that underpins the large coupling strengths which will be demonstrated below.

The CL response of the emitters positioned inside the pinholes demonstrate a remarkably different characteristics compared to those positioned outside the pinholes. The latter position we refer to as position 1 and the former as position 2.

The angle-resolved CL map acquired at Position 1 is shown in Fig. 2b, while the corresponding CL spectrum integrated over the $k_{||}$ regions marked by the dashed lines is presented in Fig. 2c. The emission consists of several weak and spectrally broad features characteristic of the uncoupled emitters and their phonon lines. Nevertheless, the integrated spectrum reveals emission centered near 443 nm, accompanied by a weak optical phonon replica at approximately 465 nm, together with broader emission bands centered near 537 nm and 688 nm, which are also recovered faintly in the EK map, as flat dispersion lines. The spectral broadening prevents the individual vibronic transitions from being clearly distinguished. This faint and broad emission response is the signature of partial quenching due to interactions with surface plasmon polaritons, particularly happening due to the fact that the hBN flake is thin, and therefore the emitters are positioned much nearer to the surface compared to those presented further below in Fig. 5.

The optical response changes when the measurement is performed directly above the plasmonic void cavity (see Fig. 2e and f, for the EK map and the integrated CL, respectively). The cavity-coupled system, consisting of the hBN flake positioned on top of the Au void cavity (Region 2 in Fig. 2a), exhibits a spectrum with well-resolved spectral features, allowing additional vibronic transitions to be identified. Besides the enhanced 443 nm B-center emission, distinct phonon sidebands become visible at 465 nm, 500 nm, and 550 nm corresponding to phonon energies of 144.3 meV, 330.6 meV, and 544.1 meV respectively[39,44]. The first sideband agrees well with the optical phonon replica reported for carbon-related B centers, whereas the second phonon sideband is consistent with a second-order two-phonon process. Half of its energy (165.3 meV) lies within the characteristic 155-200 meV optical phonon range of hBN, indicating the simultaneous emission of two optical phonons during the radiative transition[50]. The 3$^{rd}$ phonon sideband is similarly attributed to a third-order phonon scattering process.

The green emitter centered near 537 nm exhibits phonon sidebands at 554 nm and 579 nm, corresponding to phonon energies of 65.6 meV and 164.3 meV, respectively. The lower-energy replica is assigned to a localized vibrational mode associated with the defect, and the higher energy phonon could be assigned to the transverse optical (TO) phonon of hBN[51,52]. In addition, a third emission centered at 600 nm[53] exhibits a high energy optical phonon sideband at an energy of approximately 176 meV.

The broad emission around 688 nm is observed in both spectra (Fig. 2c and f). A similar ~1.8 eV emission has previously been correlated with the 4.1 eV defect emission observed in CL, suggesting different radiative transitions involving the same intra-bandgap defect state[40].

Most notably, the spectrum exhibits the emergence of a previously absent split feature centered near 550 nm, which is not observed outside the cavity. This additional peak is absent from all off-cavity measurements despite the presence of both emitters, indicating that its appearance is intrinsically linked to the plasmonic cavity rather than to the emitters alone. As discussed in the following section, this feature originates from hybridization between the vibronic sideband of the B center and the plasmonic void cavity mode, forming a new hybridized state. This fact is further revealed from our Maxwell-Bloch simulations featuring a pronounced strong coupling effect of the vibronic modes to the plasmonic cavity, as will be discussed below.

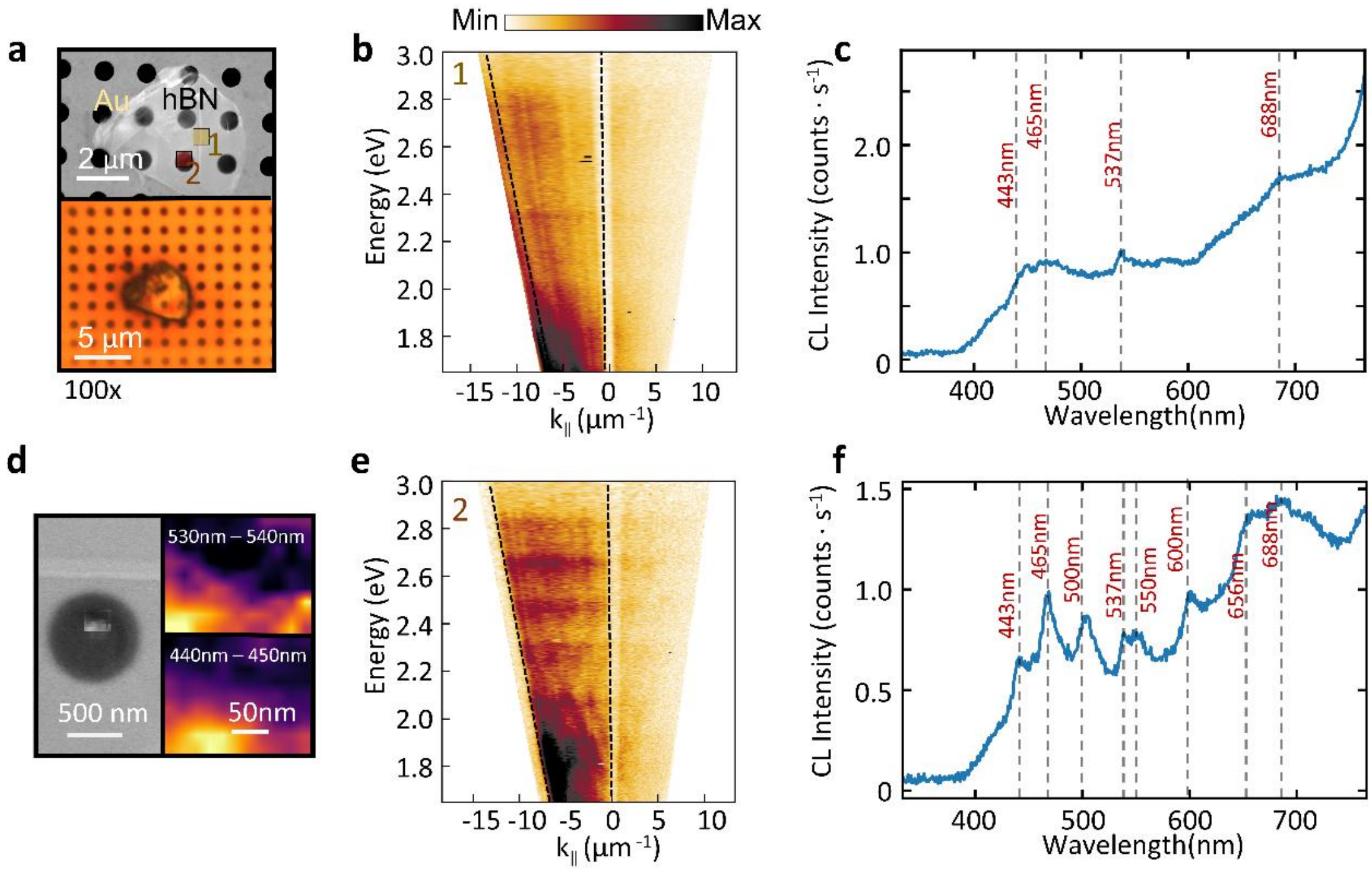

**Figure 2. CL response of uncoupled and cavity-coupled hBN emitters. (a)** SEM and optical image of the hBN flake on an Au plasmonic void cavity. Positions 1 and 2 indicate measurements outside and above the cavity, respectively. **(b,c)** Angle-resolved CL map and corresponding integrated spectrum of the uncoupled emitters acquired at position 1. **(d)** SEM image and hyperspectral CL maps demonstrating spatial distribution of the 443 nm and 537 nm emitters within the cavity. **(e,f)** Angle-resolved CL map and integrated spectrum of the cavity-coupled system acquired at position 2, revealing additionally resolved vibronic transitions and the energy split around 540 nm.

A higher-magnification SEM image of Position 2 is shown in Fig. 2d together with hyperspectral CL intensity maps integrated over the wavelength ranges 440-450 nm and 530-540 nm, corresponding to the B-center emission and the 537 nm defect, respectively. As mentioned above, the hyperspectral maps reveal that both emission bands are localized within the cavity region, confirming the coexistence of both the emitters inside the same environment. The overlapping spatial distributions indicate that both emitters are simultaneously coupled to the localized electromagnetic field of the Au void cavity. Nevertheless, only the vibronic mode of emitter 1 is strongly coupled to the cavity, whereas emitter 2 remains uncoupled, and its emission is partially quenched. The latter happens since the emission from the ZPL of emitter 2 is absorbed by the transition between 1$^{st}$ and 4$^{th}$ states of emitter 1 (See Supplementary Fig. S2 for details of the energy states corresponding to each emitter).

## Maxwell-Bloch Analysis of the Hybridized State

To further examine the origin of the additional spectral feature, the experimental observations were compared with the results of simulations obtained using a Maxwell-Bloch self-consistent numerical scheme (See Supplementary Note 1 and Supplementary Figures S1 to S4) describing the interaction between the plasmonic cavity and a network of randomly positioned emitters. The dynamics of the induced polarization by the emitters are hence modeled using a master equation coupled self-consistently to the Maxwell's equations. For results depicted in Fig. 3, we considered a number of 500 type 2 emitters positioned within the bulk of hBN as well as 400 type 1 emitters positioned near to the hBN top and bottom surfaces. The emitters were positioned randomly and at a minimal distance of 100 nm with respect to each other, with a variance of 30 nm. Interestingly. We notice, that as far as the emitter network is dense, their exact positions will not affect the results. However, considering an ensemble of considerably lower element numbers (20 type 1 emitters, and 100 type 2 emitters), the split in the energy is not observed. Hence, the collective response of emitters plays a significant role, as well as their photon-induced mutual interactions.

Figure 3a illustrates the simulated electric-field distribution of the plasmonic void cavity, highlighting the confined near field distribution as well as the electron-induced field distribution along the electron trajectory within the cavity. The induced polarization by the electron beam is the strongest contribution to the total polarization and affects the dynamics of the polarization field induced by the emitters. The fundamental mode of the cavity with a uniform radial dependence and first order resonance along the electron trajectory is centered around the wavelength of 960 nm, which is far from the emitter's ZPL and phonon sidebands (Fig. 3b). Higher-order cavity resonances occur in the visible spectral range, consistent with the experimentally measured optical response of the hBN–Au structure (See Supplementary Note

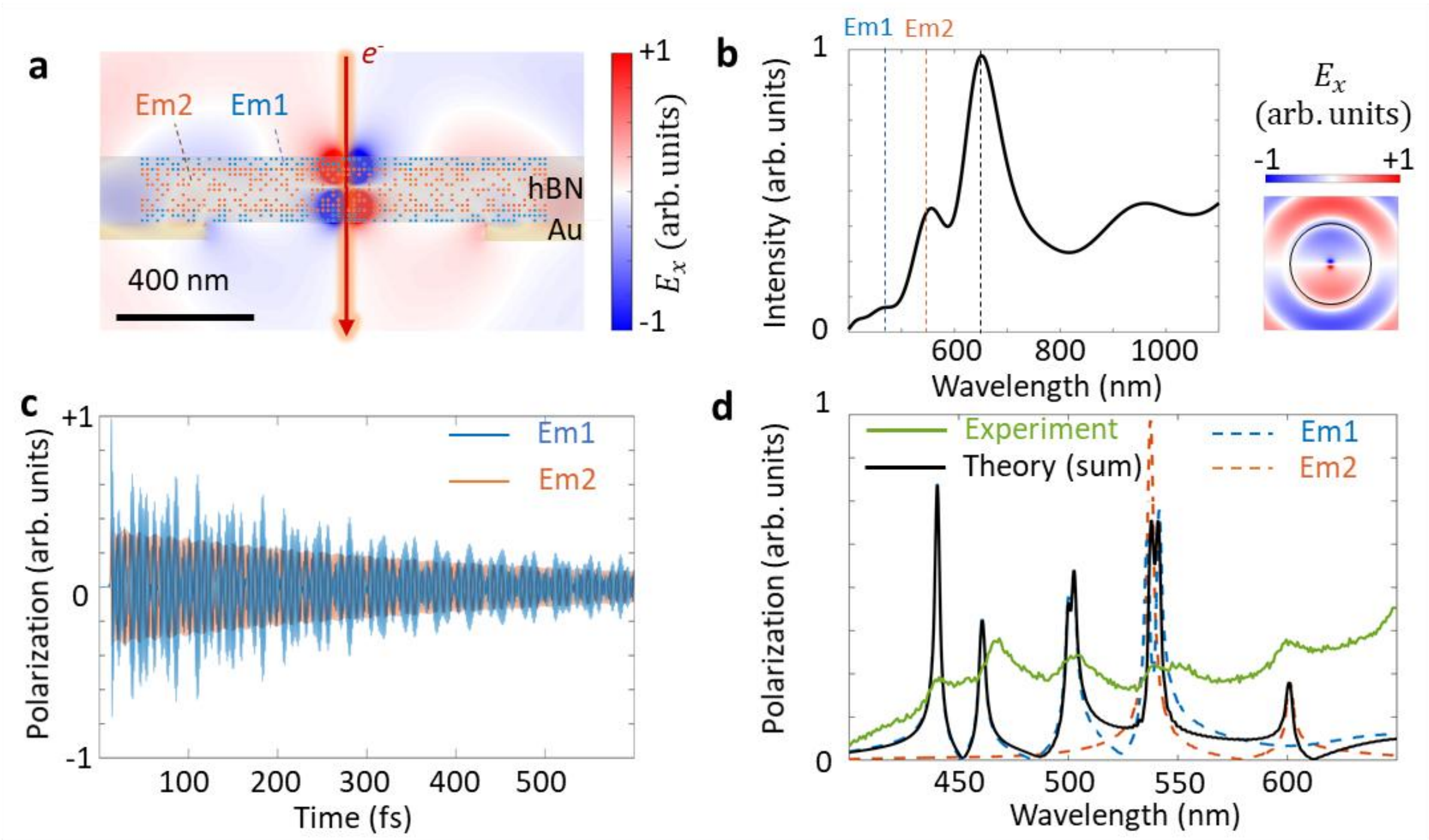

**Figure 3. Maxwell-Bloch simulations of cavity-mediated emitter coupling. (a)** Simulated electron-induced near-field distribution of the in-plane component of the electric field of the plasmonic void cavity in the absence of both emitters at the wavelength of 537 nm. **(b)** Calculated cavity resonance spectrum with the positions of emitter 1 (Em1) and emitter 2 (Em2) ZPLs peaks marked with blue and orange dashed lines, respectively. Moreover, the position of the fundamental mode of the cavity, which is a photonic mode confined in the hole is marked by blue dashed line. **(c)** Temporal evolution of the individual emitter polarizations, for two emitters positioned near to the rim of the cavity. **(d)** Comparison between the simulated (black) and experimental (green) spectra. Spectra of individual emitter polarizations are displayed by blue and orange dashed lines for Em1 and Em2, respectively. The demonstrated spectra are averaged over each emitter type (for Em1 and Em2) and over the entire network of emitters (for the black line).

S7). In particular, the absorption spectrum calculated from the measured reflectance and transmittance exhibits a broad resonance in this spectral region (Supplementary Fig. S14). The 2$^{nd}$ and 3$^{rd}$ order resonances are centered around 650 nm and 537 nm, where the latter is perfectly tuned with the ZPL of emitter 2 and the 4$^{th}$ order vibrational mode of the emitter 1 (See Supplementary Fig S1 for the field distributions associated with these modes).

The interaction of the fast-moving electron (kinetic energy of 30 keV) with the structure takes place within a time duration of only 2.43 fs – however, the polarization associated with the emitters last for as long as 500 fs (Fig. 3c). The radiative decay rates of the emitters and their phonon sidebands are taken to be from 13 meV to 66 meV in the simulations, in close agreement with the experimentally observed linewidths. These long dynamics substantiate a coherent and resonant transfer of energy between the emitter network and the plasmonic cavity, modulating the polarization as well as the density matrix elements (Fig. 3c and Supplementary Fig. S2 and S3). Particularly, the populations of the electrons in the 3$^{rd}$ and 4$^{th}$ order vibrational states of type 1 emitter are strongly modulated, and the coherence terms of the density matrix elements for type 1 emitters last for substantially longer time to mediate the long-lasting mutual interactions between the emitter polarizations and the plasmonic field (Supplementary Fig. S2 and S3). Therefore, the theoretically observed energy split is associated with the 4$^{th}$ vibrational mode of type 1 emitter (as well as its 3$^{rd}$ vibrational mode), whereas the ZPL line associated with emitter 2 does not demonstrate any energy split. The Fourier-transformed polarizations for each emitter sets, summed over the entire set of emitters, as well as the summation of both polarizations as well, demonstrate a theoretical energy split at the order of 23.6 meV, slightly lower than the experimentally observed value of 36.24 meV for the CL spectrum. This fact is associated with the higher excitation yield of the emitters in the experiments, where secondary and back-scattered electrons could further excite the emitters and substantiate a larger probability of exciting the emitters per electron excitations. However, we observe a quite similar values for the energy split for our simulations, where light pulses are used (Supplementary Fig. S4), providing evidence that the interactions between the emitter networks and the plasmonic cavity substantiate the dynamics of energy exchange, rather than the excitation scheme. Moreover, whereas the positions of the peaks correspond well to the experimentally acquired values (compare black and green lines in Fig. 3d), the experimental CL spectra manifest a significantly larger broadening. This is however attributed mainly to the fact that the theoretical results are the induced polarization density, and not the field propagating to the far-field.

**Phonon-Assisted Optical Excitation**

In addition to our Maxwell-Bloch simulations, several possible origins of the additional 550nm split spectral feature were examined using different experimental features and approaches. Its assignment to an independent defect can be excluded because no corresponding emission was detected outside the cavity or in regions containing only one emitter. The feature is also inconsistent with a conventional phonon replica, as both its spectral position and linewidth differ from the vibronic sidebands measured for the uncoupled emitters. To determine whether the split depends on the excitation mechanism, PL measurements were performed on the same positions as the CL measurements. Unlike CL, which generates electron-hole pairs over a broad energy distribution, PL selectively excites the sample with photons of well-defined energies. Excitation wavelengths between 460 and 485 nm are used for measurements, by tuning the excitation wavelength using a super-continuum laser source combined with a tunable filter (see Methods Sections). This energy range is sufficient to excite the 537 nm defect but lies at lower energy than the 443 nm zero-phonon transition. Despite the absence of direct resonant excitation, the PL spectra exhibit the same additional split observed under CL, while the ZPL of Emitter 2 remains unchanged.

The persistence of this feature under both electron-beam and optical excitation demonstrates that its formation is independent of the excitation mechanism and cannot be attributed to an excitation-specific artifact. The PL emission map shown in Fig. 4a exhibits two distinct responses. A Stokes-shifted emission band continuously follows the excitation wavelength with an energy shift of ~0.17 eV. This energy overlaps the Reststrahlen band of the infrared-active $E_{1u}$ LO-TO phonon pair and lies close to the first-order $E_{2g}$ phonon of hBN[54,55]. Defect-induced inversion-symmetry breaking relaxes the optical selection rules, allowing the $E_{1u}$ phonons to contribute to the phonon-assisted relaxation process[27]. The Stokes-shifted emission exhibits two partially resolved components with energy shifts of approximately 0.17 and 0.20 eV. Remarkably, a third Stokes-shifted feature with energy shift of approximately 0.37 eV is also observed. The Au substrate enhances the weak Raman process through surface-enhanced Raman scattering (SERS)[56], resulting in phonon-assisted excitation of the coupled state. The corresponding phonon assignments and calculated Stokes energies are presented in Supplementary Note 2 and Supplementary Fig. S5. The PL spectra individually normalized to the measured excitation laser intensity, show the Stokes-shifted emission traversing the cavity-splitting spectral range while the split mode remains pinned at approximately 544 nm throughout the excitation range (Fig. 4b-d). This excitation mechanism inferred from these measurements is therefore a coherent excitation of defects via the down-converted peaks due to the Raman scattering, followed by energy transfer between emitter 1 and emitter 2 sets (see the schematics in Fig. 4e).

The excitation efficiency is quantified by the defect intensity ratio defined as the integrated intensity of the hybridized spectral region (536nm-548nm) normalized to the total PL intensity (Fig. 4f). The ratio exhibits a pronounced maximum for excitation at 465 nm. At this excitation wavelength, the corresponding Stokes-shifted emission is centered near 540 nm, spectrally overlapping with the 537 nm defect and thereby maximizing its excitation efficiency. Simultaneously, the same phonon-assisted excitation pathway efficiently populates the vibronic state of the 443 nm emitter, enabling its hybridization with the cavity mode discussed above in the theory section.

This correspondence identifies the vibronic sideband as the dominant excitation channel for the coupled state and demonstrates that population of the vibronic level alone is sufficient to establish the emitter-cavity interaction under optical excitation. Supplementary Note 3 compares the hybridized-mode splitting measured under CL and PL and presents the corresponding energy splitting at $\delta E =$36.24meV and 24.09meV. Despite a lower splitting energy under optical excitation, both measurements consistently exhibit the same cavity-induced hybridized state. Our numerical results using the Maxwell-Bloch frameworks excellently reproduce the energy split, when both light resonantly excite the emitters, or first excite a certain vibrational state of emitter 1 and sequentially interacts with the cavity (See Supplementary Note 1 and Supplementary Fig. S4). In fact, the numerically-obtained energy split of 24.5 meV, is in an excellent agreement with our experimental observations.

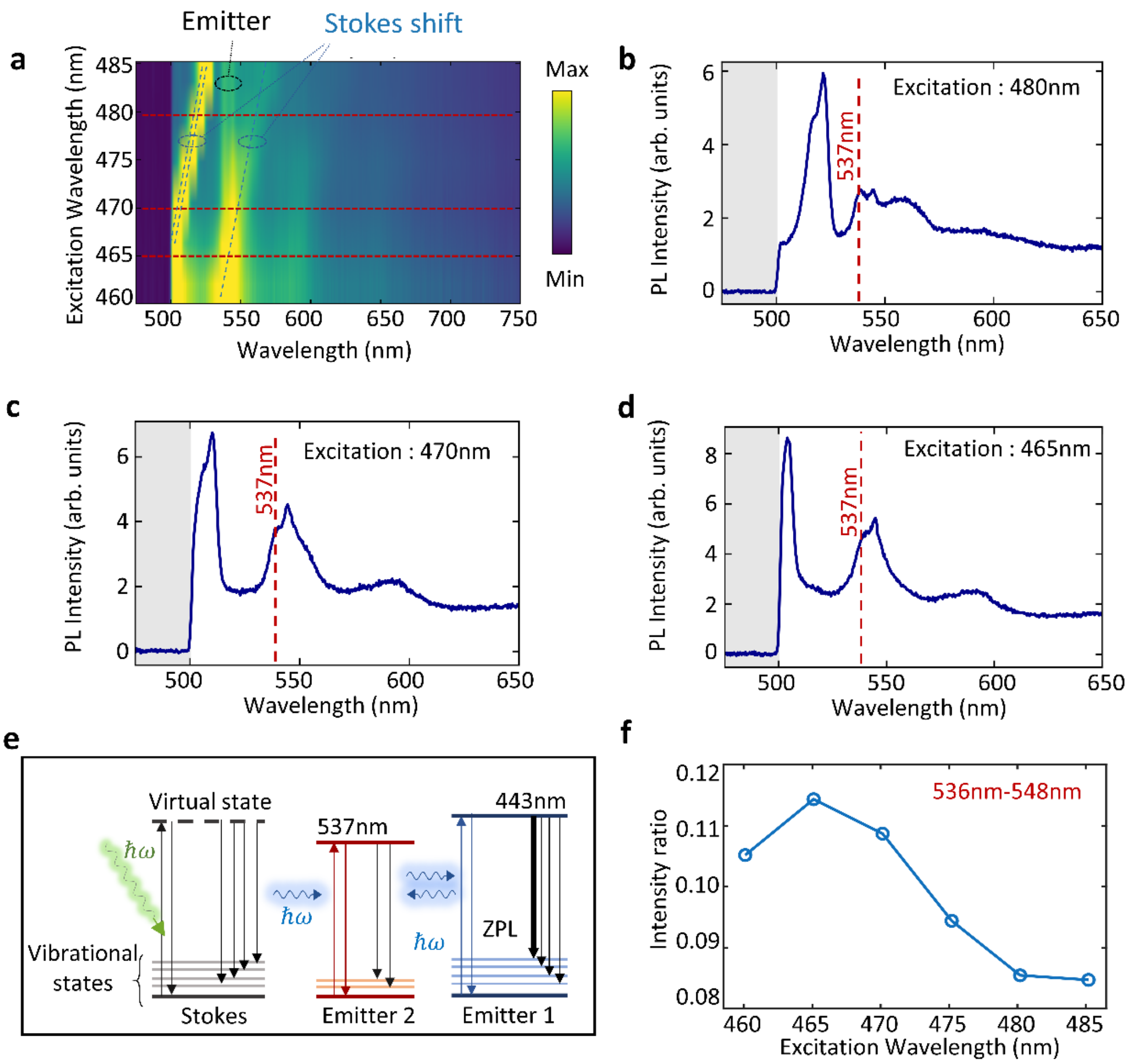

**Figure 4. Phonon-assisted optical excitation of the cavity-coupled hybridized state. (a)** Excitation-dependent PL map of the coupled emitter-cavity system, normalized individually to the laser intensity. The Stokes-shifted emission follows the excitation wavelength, while the hybridized emission remains limited to the same wavelengths. **(b-d)** Representative PL spectra for excitation wavelengths of 480, 470, and 465 nm, respectively, showing the persistence of the cavity-induced split feature under all excitations. The dashed line marks the 537 nm ZPL of Emitter 2. **(e)** Schematic of the phonon-assisted excitation pathway. Excitation below the 443 nm ZPL of Emitter 1 populates its vibronic level, enabling Stokes-assisted excitation of the cavity-coupled hybridized state. **(f)** Defect Intensity ratio of the hybridized spectral region as a function of excitation wavelength, exhibiting a maximum near the first phonon replica of Emitter 1 at 465nm.

The PL measurements additionally reveal the localized nature of the coupling process. Consequently, the hybridized spectral feature is observed only when the laser is focused directly onto the cavity and is absent for neighboring excitation positions. This positional dependence further confirms the localized nature of the cavity-mediated interaction. A detailed analysis of the spatial sensitivity is provided in Supplementary Note 4.

## Collective Excitations

The cavity–emitter hybridization observed here cannot be accounted for by the coupling of an individual emitter to the plasmonic cavity mode. From the pseudo-Voigt analysis described above, the linewidth associated with the fourth vibronic transition of the type-1 emitter is determined to be $\gamma$ = 60 meV. This linewidth substantially exceeds the coupling strength $g_1$ expected for an individual emitter interacting with the cavity mode (see Supplementary Note 1), placing the single-emitter system outside the resolved strong-coupling regime.

The situation changes fundamentally when an ensemble of emitters couples coherently to the same electromagnetic mode. For $N$ equivalent emitters, the optically bright collective state couples to the cavity with the collectively enhanced coupling strength $\sqrt{N}g_1$ whereas the intrinsic damping of an individual emitter does not acquire the same enhancement in the absence of collective dissipative effects. Consequently, the strong-coupling condition can be expressed in terms of the collective coupling as $\sqrt{N}g_1 > (\gamma+\kappa)/4$, with $\kappa$ and $\gamma$ denoting the cavity and emitter spectral FWHM linewidths, respectively. Thus, an ensemble can enter the collective strong-coupling regime even when the corresponding single-emitter coupling is insufficient to overcome the dissipative losses.[57-60]

Experimentally, the two hybridized branches observed in CL are separated by 36.24 meV. Importantly, this splitting emerges under electron-beam excitation, which simultaneously drives a spatially distributed ensemble of emitters coupled to the same plasmonic mode. The observation is therefore consistent with a collectively enhanced light–matter interaction rather than independent coupling of isolated emitters. This interpretation is further supported by the self-consistent Maxwell–Bloch calculations (see Supplementary Note S1), in which the experimentally observed spectral splitting emerges only when a sufficiently large ensemble of emitters participates in the interaction, while it progressively collapses upon reducing the number of coupled emitters.

**Thickness and EHT Dependence**

The measured optical response is strongly influenced by the local morphology and thickness of the hBN flake, which determine both the electron interaction volume and the relative contributions of defect and background emission. To investigate these effects while avoiding cavity interactions, we next perform CL measurements on hBN flakes deposited on a uniformly Au-coated glass substrate. The flake thickness is first characterized using a MarSurf CM explorer (See Methods Section), and measurement positions are selected on regions with different thicknesses (Fig. 5a and b). The CL measurements are then performed while systematically varying the electron-beam accelerating voltage (EHT) from 8 to 20 kV. The thin flake has a maximum thickness of approximately 700 nm, whereas the thick region reaches about 4 µm.

Increasing the accelerating voltage increases both the penetration depth of the primary electrons and the interaction volume within the crystal, thereby modifying the excitation probability of defects located at different depths. This provides a direct experimental probe of the depth-dependent spatial positioning of the emitters within the flake. The emission associated with Emitter 2 (537 nm) increases progressively versus acceleration voltage, particularly in the thin flake, where it reaches its highest CL intensity at an EHT of 20 kV. This trend is opposite to that observed for Emitter 1, whose relative intensity decreases with increasing penetration depth and is most pronounced in the at low accelerating voltages.

This indicates distinct depth distributions of the defects within the crystal. Emitter 2 is therefore consistent with a defect population distributed throughout the flake volume, whereas Emitter 1 exhibits the characteristics of a near-surface defect population, generated during the liquid exfoliation process near to the surfaces.

The present measurements hence indicate that the two emitters originate from physically distinct defect environments. This trend is less clear for the thick flake. Emitter 2 reaches its highest intensity at an accelerating voltage of 20 kV, whereas Emitter 1 exhibits its maximum intensity at 14 kV. However, Emitter 1 remains visible over the entire voltage range, including 8 kV, while Emitter 2 is only observed at higher accelerating voltages. The different behavior could be influenced by the stronger background emission in the thick flakes. A detailed analysis on the effect of the morphology on the optical responses of hBN is presented in the Supplementary Notes 5, which compares the PL and CL spectra of flakes and show that increasing flake thickness progressively increases the broadband background contribution, thereby reducing the relative visibility of the defect emission.

As Emitter 2 has been attributed to dislocations or carbon-related defect complexes that are distributed throughout the flake, an efficient excitation scheme requires electrons to penetrate deeper into the material, leading to a stronger signal at higher accelerating voltages. In addition, this sample also exhibits substrate-assisted excitation, allowing the defect emission to be detected in CL even when the electron beam is positioned outside the flake[56]. The latter is due to both the excitation of surface plasmon polaritons that sequentially interact with the emitter, as well as the secondary and back scattered electrons from the substrate that can reach the flake.

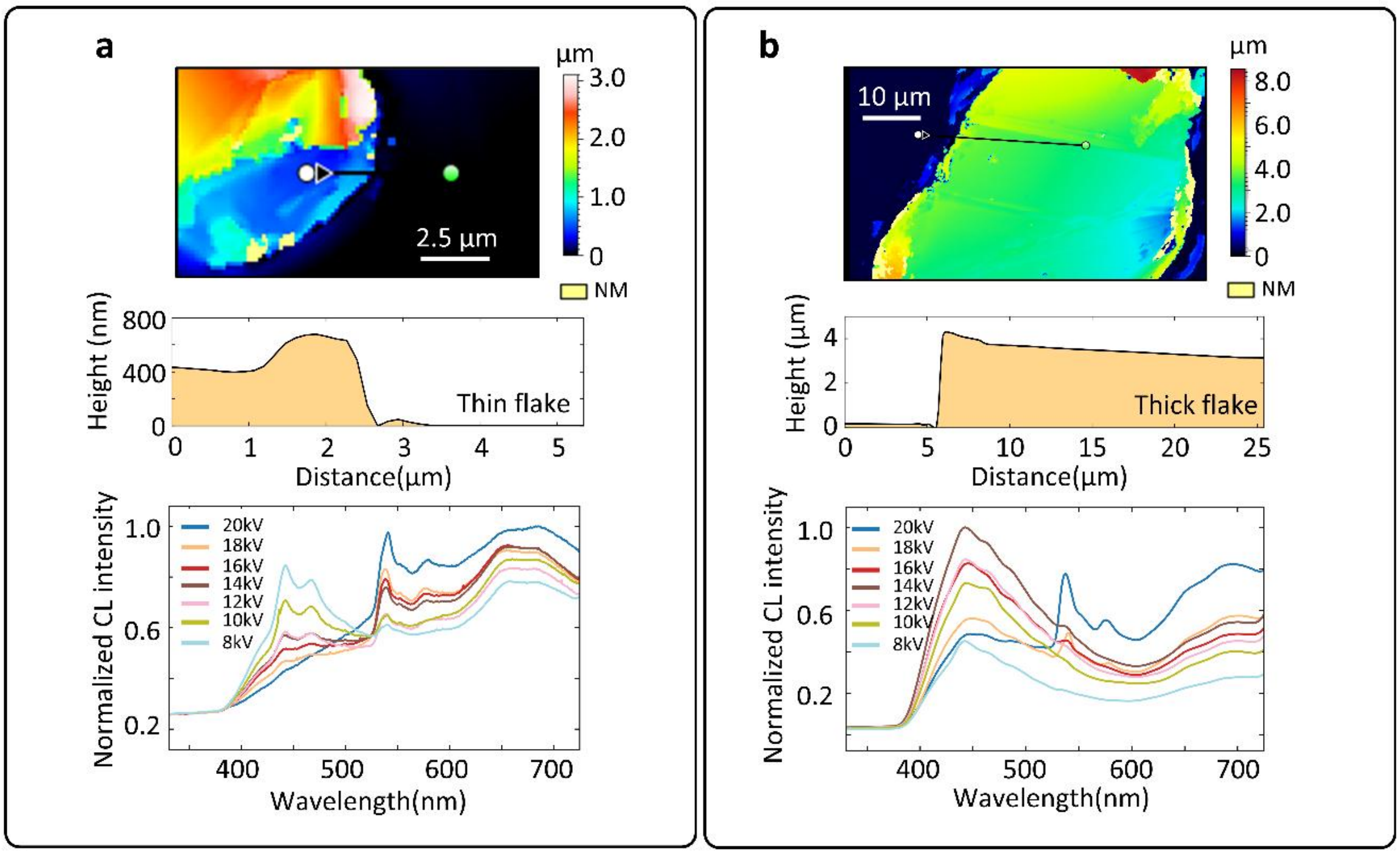

**Figure 5. Thickness-dependent CL of hBN flakes. (a,b)** MarSurf thickness maps (top), corresponding height profiles (middle), and normalized CL spectra (bottom) of representative thin and thick hBN flakes measured at accelerating voltages from 8 to 20 kV, and the beam current of 8 nA.

The fact that the observed strong-coupling effect is attributed to the $4^{th}$ vibrational state of emitter 1 rather than the ZPL of emitter 2, is further confirmed by these measurements. Emitter 1 is generated during the exfoliation process and is located near to the surfaces of the hBN flake, hence couples significantly better to the evanescent tails of the plasmonic cavity penetrating into the hBN flake positioned on top of the cavity.

Moreover, for a vibronic transition, the effective transition strength is weighted by its Franck–Condon factor.[61] Consequently, different vibronic transitions of a given color center can exhibit substantially different coupling efficiencies to the same electromagnetic mode. Indeed, phonon-assisted optical transitions in solid-state cavity-quantum electrodynamics systems have been shown to efficiently mediate energy transfer between quantum emitters and cavity photons, demonstrating that lattice vibrations can provide active coupling channels rather than acting exclusively as a source of decoherence.[7]

## Discussions

The combined experimental and theoretical results identify a previously unexplored cavity-mediated interaction in hBN. The plasmonic void cavity strongly interacts with the vibronic transition of the high-energy 443nm blue emitter while the spectrally overlapping lower-energy emitter remains in the weak-coupling regime. Agreement between CL, excitation-dependent PL, and Maxwell-Bloch simulations demonstrates that the observed mode splitting originates from hybridization with the vibronic level of Emitter 1 rather than the ZPL of emitter 2. The persistence of the split mode under excitation energies below the 443 nm electronic transition further establishes that phonon-assisted excitation of the vibronic level is sufficient to access the hybridized state.

Our observations demonstrate that electron-phonon interactions can actively mediate coherent emitter-cavity coupling rather than acting solely as a source of decoherence. By using vibronic transitions as an active pathway of the coupling process, the present approach extends cavity quantum electrodynamics beyond direct ZPL interactions and introduces an efficient route for controlling light-matter coupling. This mechanism provides new opportunities for phonon-assisted strong coupling, cavity-mediated interactions between spectrally distinct emitters, and the rational design of hybrid quantum photonic architectures based on layered materials.

## Sample Fabrication

Hexagonal boron nitride (hBN) crystals were obtained from HQ Graphene and exfoliated into thin nanosheets by liquid-phase exfoliation in isopropanol (≥99.8%, Merck). Bulk hBN crystals were ultrasonicated (320 W, Bandelin Sonorex RK100H) in an ice bath using a 5 min on / 1 min off cycle for a total sonication time of 180 min to minimize heating and solvent evaporation. The resulting dispersion was drop-cast onto commercially available holey Au TEM grids (Quantifoil) to form plasmonic void cavity structures, where suspended hBN flakes bridged the circular holes. The samples were dried under ambient conditions prior to optical and electron-beam characterization.

## CL Spectroscopy

CL measurements were performed using a ZEISS Sigma field-emission scanning electron microscope equipped with a Delmic SPARC CL detection system[62-66]. Unless otherwise stated, spectra were acquired at an accelerating voltage of 20 kV. CL emission was collected using an aluminum-coated parabolic mirror positioned on top of the sample and directed to a CCD-equipped spectrometer. Hyperspectral CL maps were recorded with pixel dwell times ranging from 250 to 800 ms depending on the measurement. Angle-resolved energy-momentum (E–k) spectroscopy was performed using a 200 μm entrance slit with an acquisition time of 180 s. Additional measurements at accelerating voltages between 8 and 20 kV were carried out to investigate the influence of electron penetration depth on the excitation of the different emitters.

## PL Spectroscopy

PL measurements were performed using a supercontinuum laser source (SC-OEM, YSL Photonics) combined with an acousto-optic tunable filter providing an excitation bandwidth of approximately 2 nm. The laser operated at a repetition rate of 5 MHz with pulse durations below 100 ps. The Nikon ECLIPSE Ti2 inverted microscope is used with a Princeton Instruments SpectraPro HRS-500 spectrometer (NT&C Company)[67]. A 100× objective (NA = 0.9) was used to focus the excitation onto individual hBN flakes above plasmonic void cavities. The excitation power was controlled using neutral-density filters. The emitted PL was separated from the excitation light using dichroic mirrors together with long-pass and short-pass filters. Standard measurements were acquired using a 150 grooves $mm^{-1}$ grating. The excitation laser intensity was independently measured for each excitation wavelength using an AvaSpec-ULS2048CL-EVO spectrometer (Avantes). All measurements were performed at room temperature.

## Thickness Measurements

The three-dimensional surface topography and flake thickness were measured using a MarSurf3D CM confocal microscope (Mahr GmbH, Germany). The system employs Confocal Multi-Pinhole (CMP) technology, in which illumination is focused onto the sample through a rotating multi-pinhole disc and the reflected light is spatially filtered to suppress contributions from adjacent measurement points. The resulting confocal response is used to determine the height at each pixel, providing a three-dimensional surface profile. Flake thicknesses were extracted from the measured height difference between the hBN surface and the surrounding substrate. The resolution for extracting the thickness is approximately 2 nm[68].

## Numerical Modelling

The interaction between the plasmonic void cavity and the two hBN emitters was investigated using a Maxwell-Bloch formalism. Full descriptions of the implemented code and supporting results are provided in the Supplementary Note 1 and Supplementary Fig. S1 to S4.

## Acknowledgements

This Project received funding from the Volkswagen Foundation (Momentum Grant), European Research Council (ERC Consolidator Grant UltraSpecT with no. 101170341; ERC Proof-of-Concept Grant UltraCoherentCL with no. 101157312) and from Deutsche Forschungsgemeinschaft (Project Nr. 525347396).

## Data availability

The data that support the findings of this study are available from the corresponding author upon reasonable request.

## Code availability

The analytical code used to simulate the data is available from the corresponding author on request.

## Competing interests

The authors declare no competing interests.

## Contributions

M.T. and B.G. performed the sample synthesis. B.G. performed the CL spectroscopy with supports from M.T. P.S. performed the PL measurements. B.G. and P.S. wrote the manuscript together with M.T. and N.T. All authors contributed to the data analysis and visualization. N.T. performed the Maxwell-Bloch simulations and supervised the work.

## References

1 Evans, R. E. *et al.* Photon-mediated interactions between quantum emitters in a diamond nanocavity. *Science* **362**, 662-665 (2018). https://doi.org/doi:10.1126/science.aau4691
2 Gaur, K. *et al.* Scalable quantum photonic platform based on site-controlled quantum dots coupled to circular Bragg grating resonators. *Light: Science & Applications* **15**, 260 (2026). https://doi.org/10.1038/s41377-026-02343-0
3 Jelezko, F., Gaebel, T., Popa, I., Gruber, A. & Wrachtrup, J. Observation of Coherent Oscillations in a Single Electron Spin. *Physical Review Letters* **92**, 076401 (2004). https://doi.org/10.1103/PhysRevLett.92.076401
4 Kurtsiefer, C., Mayer, S., Zarda, P. & Weinfurter, H. Stable Solid-State Source of Single Photons. *Physical Review Letters* **85**, 290-293 (2000). https://doi.org/10.1103/PhysRevLett.85.290
5 Michler, P. *et al.* A Quantum Dot Single-Photon Turnstile Device. *Science* **290**, 2282-2285 (2000). https://doi.org/doi:10.1126/science.290.5500.2282
6 Yoshie, T. *et al.* Vacuum Rabi splitting with a single quantum dot in a photonic crystal nanocavity. *Nature* **432**, 200-203 (2004). https://doi.org/10.1038/nature03119
7 Hohenester, U. Cavity quantum electrodynamics with semiconductor quantum dots: Role of phonon-assisted cavity feeding. *Physical Review B* **81**, 155303 (2010). https://doi.org/10.1103/PhysRevB.81.155303
8 Ohta, R. *et al.* Strong coupling between a photonic crystal nanobeam cavity and a single quantum dot. *Applied Physics Letters* **98**, 173104 (2011). https://doi.org/10.1063/1.3579535
9 Reithmaier, J. P. *et al.* Strong coupling in a single quantum dot-semiconductor microcavity system. *Nature* **432**, 197-200 (2004). https://doi.org/10.1038/nature02969
10 Wilson-Rae, I. & Imamoğlu, A. Quantum dot cavity-QED in the presence of strong electron-phonon interactions. *Physical Review B* **65**, 235311 (2002). https://doi.org/10.1103/PhysRevB.65.235311
11 Albert, F. *et al.* Microcavity controlled coupling of excitonic qubits. *Nature Communications* **4**, 1747 (2013). https://doi.org/10.1038/ncomms2764

12 Kim, H., Sridharan, D., Shen, T. C., Solomon, G. S. & Waks, E. Strong coupling between two quantum dots and a photonic crystal cavity using magnetic field tuning. *Optics Express* **19**, 2589-2598 (2011). https://doi.org/10.1364/oe.19.002589

13 Laucht, A. *et al.* Mutual coupling of two semiconductor quantum dots via an optical nanocavity. *Physical Review B* **82**, 075305 (2010). https://doi.org/10.1103/PhysRevB.82.075305

14 Fink, J. M. *et al.* Dressed Collective Qubit States and the Tavis-Cummings Model in Circuit QED. *Physical Review Letters* **103**, 083601 (2009). https://doi.org/10.1103/PhysRevLett.103.083601

15 Badolato, A. *et al.* Deterministic Coupling of Single Quantum Dots to Single Nanocavity Modes. *Science* **308**, 1158-1161 (2005). https://doi.org/10.1126/science.1109815

16 Hu, S. *et al.* Robust consistent single quantum dot strong coupling in plasmonic nanocavities. *Nature Communications* **15**, 6835 (2024). https://doi.org/10.1038/s41467-024-51170-7

17 Sapienza, L., Davanço, M., Badolato, A. & Srinivasan, K. Nanoscale optical positioning of single quantum dots for bright and pure single-photon emission. *Nature Communications* **6**, 7833 (2015). https://doi.org/10.1038/ncomms8833

18 Senellart, P., Solomon, G. & White, A. High-performance semiconductor quantum-dot single-photon sources. *Nature Nanotechnology* **12**, 1026-1039 (2017). https://doi.org/10.1038/nnano.2017.218

19 Aharonovich, I., Englund, D. & Toth, M. Solid-state single-photon emitters. *Nature Photonics* **10**, 631-641 (2016). https://doi.org/10.1038/nphoton.2016.186

20 Lodahl, P., Mahmoodian, S. & Stobbe, S. Interfacing single photons and single quantum dots with photonic nanostructures. *Reviews of Modern Physics* **87**, 347-400 (2015). https://doi.org/10.1103/RevModPhys.87.347

21 Kianinia, M. *et al.* Robust Solid-State Quantum System Operating at 800 K. *ACS Photonics* **4**, 768-773 (2017). https://doi.org/10.1021/acsphotonics.7b00086

22 Kim, S. H. Color Centers in Hexagonal Boron Nitride. *Nanomaterials* **13**, 2344 (2023). https://doi.org/10.3390/nano13162344

23 Caldwell, J. D. Photonics with hexagonal boron nitride. *Nature Reviews Materials* **4**, 552-567 (2019). https://doi.org/10.1038/s41578-019-0124-1

24 Wigger, D. Phonon-assisted emission and absorption of individual color centers in hexagonal boron nitride. *2D Materials* **6** (2019).

25 Jungwirth, N. R. & Fuchs, G. D. Optical Absorption and Emission Mechanisms of Single Defects in Hexagonal Boron Nitride. *Physical Review Letters* **119**, 057401 (2017). https://doi.org/10.1103/PhysRevLett.119.057401

26 Preuß, J. A. *et al.* Resonant and phonon-assisted ultrafast coherent control of a single hBN color center. *Optica* **9**, 522-531 (2022).

27 Singh, P., Garg, B., Black, M., Taleb, M. & Talebi, N. Defect-mediated inelastic light scattering by infrared phonons in hexagonal boron nitride. *Advanced Functional Materials* (2026).

28 Majumdar, A., Bajcsy, M., Rundquist, A., Kim, E. & Vučković, J. Phonon-mediated coupling between quantum dots through an off-resonant microcavity. *Physical Review B* **85**, 195301 (2012). https://doi.org/10.1103/PhysRevB.85.195301

29 Grosso, G. *et al.* Tunable and high-purity room temperature single-photon emission from atomic defects in hexagonal boron nitride. *Nature Communications* **8**, 705 (2017). https://doi.org/10.1038/s41467-017-00810-2

30 Hayee, F. *et al.* Revealing multiple classes of stable quantum emitters in hexagonal boron nitride with correlated optical and electron microscopy. *Nature Materials* **19**, 534-539 (2020). https://doi.org/10.1038/s41563-020-0616-9

31 Shotan, Z. Photoinduced Modification of Single-Photon Emitters in Hexagonal Boron Nitride. *ACS Photonics* **3**, 2490-2496 (2016). https://doi.org/10.1021/acsphotonics.6b00736

32 Taleb, M. *et al.* Ultrafast phonon-mediated dephasing of color centers in hexagonal boron nitride probed by electron beams. *Nature Communications* **16**, 2326 (2025). https://doi.org/10.1038/s41467-025-57584-1

33 Bourrellier, R. *et al.* Bright UV Single Photon Emission at Point Defects in h-BN. *Nano Letters* **16**, 4317-4321 (2016). https://doi.org/10.1021/acs.nanolett.6b01368
34 Fröch, J. E. *et al.* Coupling Hexagonal Boron Nitride Quantum Emitters to Photonic Crystal Cavities. *ACS Nano* **14**, 7085-7091 (2020). https://doi.org/10.1021/acsnano.0c01818
35 Häußler, S. Tunable Fiber-Cavity Enhanced Photon Emission from Defect Centers in hBN. *Advanced Optical Materials* **9**, 2002218 (2021). https://doi.org/10.1002/adom.202002218
36 Vogl, T., Lecamwasam, R., Buchler, B. C., Lu, Y. & Lam, P. K. Compact Cavity-Enhanced Single-Photon Generation with Hexagonal Boron Nitride. *ACS Photonics* **6**, 1955-1962 (2019). https://doi.org/10.1021/acsphotonics.9b00314
37 Genc, S. *et al.* Disorder-Engineered Hybrid Plasmonic Cavities for Emission Control of Defects in hBN. *ACS Photonics* **13**, 937-948 (2026). https://doi.org/10.1021/acsphotonics.5c02063
38 Tauro, S. M. Photophysics of cavity-coupled quantum emitters in hexagonal boron nitride. *Physical Review B* **113** (2026).
39 Hou, H., Hua, M. & Kolluru, V. S. C. Nanometer Resolution Structure-Emission Correlation of Individual Quantum Emitters via Enhanced Cathodoluminescence in Twisted Hexagonal Boron Nitride. *Advanced Materials* **37**, e01611 (2025). https://doi.org/10.1002/adma.202501611
40 Ciampalini, G. *et al.* Light emission properties of mechanical exfoliation induced extended defects in hexagonal boron nitride flakes. *2D Materials* **9**, 035018 (2022). https://doi.org/10.1088/2053-1583/ac6f09
41 Tran, T. T., Wang, D., Xu, Z.-Q. & Yang, A. Deterministic Coupling of Quantum Emitters in 2D Materials to Plasmonic Nanocavity Arrays. *Nano Letters* **17**, 2634-2639 (2017). https://doi.org/10.1021/acs.nanolett.7b00444
42 Plo, J. Isotope Substitution and Polytype Control for Point Defects Identification: The Case of the Ultraviolet Color Center in Hexagonal Boron Nitride. *Physical Review X* **15**, 021045 (2025). https://doi.org/10.1103/PhysRevX.15.021045
43 Gérard, D. *et al.* Resonance fluorescence and indistinguishable photons from a coherently driven B centre in hBN. *Nature Communications* **17**, 1843 (2026). https://doi.org/10.1038/s41467-026-68555-5
44 Núñez Marcos, A. s. *et al.* Deterministic Generation of Single B Centers in hBN by One-to-One Conversion from UV Centers. *ACS Nano* **20**, 10404-10411 (2026). https://doi.org/10.1021/acsnano.5c19102
45 Proscia, N. V. *et al.* Near-deterministic activation of room-temperature quantum emitters in hexagonal boron nitride. *Optica* **5**, 1128 (2018). https://doi.org/10.1364/optica.5.001128
46 Koperski, M. *et al.* Midgap radiative centers in carbon-enriched hexagonal boron nitride. *Proceedings of the National Academy of Sciences* **117**, 13214-13219 (2020). https://doi.org/doi:10.1073/pnas.2003895117
47 Xu, Z.-Q. *et al.* Single photon emission from plasma treated 2D hexagonal boron nitride. *Nanoscale* **10**, 7957-7965 (2018). https://doi.org/10.1039/c7nr08222c
48 Ida, T., Ando, M. & Toraya, H. Extended pseudo-Voigt function for approximating the Voigt profile. *Journal of Applied Crystallography* **33**, 1311-1316 (2000). https://doi.org/10.1107/s0021889800010219
49 Bittorf, P. H., Davoodi, F., Taleb, M. & Talebi, N. Mapping optical Bloch modes of a plasmonic square lattice in real and reciprocal spaces using cathodoluminescence spectroscopy. *Optics Express* **29**, 34328-34340 (2021). https://doi.org/10.1364/OE.437984
50 Mendelson, N. *et al.* Identifying carbon as the source of visible single-photon emission from hexagonal boron nitride. *Nature Materials* **20**, 321-328 (2021). https://doi.org/10.1038/s41563-020-00850-y
51 Arı, O., Polat, N., Fırat, V., Çakır, Ö. & Ateş, S. Temperature-Dependent Spectral Properties of Hexagonal Boron Nitride Color Centers. *ACS Photonics* **12**, 1676-1682 (2025). https://doi.org/10.1021/acsphotonics.4c02616
52 Kang, T. *et al.* Ultrafast nonlinear phonon response of few-layer hexagonal boron nitride. *Physical Review B* **104**, L140302 (2021). https://doi.org/10.1103/PhysRevB.104.L140302

53 Tang, T. W. *et al.* Structured-Defect Engineering of Hexagonal Boron Nitride for Identified Visible Single-Photon Emitters. *ACS Nano* **19**, 8509-8519 (2025). https://doi.org/10.1021/acsnano.4c11413
54 Caldwell, J. D. *et al.* Sub-diffractional volume-confined polaritons in the natural hyperbolic material hexagonal boron nitride. *Nature Communications* **5**, 5221 (2014). https://doi.org/10.1038/ncomms6221
55 Geick, R., Perry, C. H. & Rupprecht, G. Normal Modes in Hexagonal Boron Nitride. *Physical Review* **146**, 543-547 (1966). https://doi.org/10.1103/PhysRev.146.543
56 Tong, L., Zhu, T. & Liu, Z. Approaching the electromagnetic mechanism of surface-enhanced Raman scattering: from self-assembled arrays to individual gold nanoparticles. *Chemical Society Reviews* **40**, 1296-1304 (2011). https://doi.org/10.1039/c001054p
57 Blaha, M., Johnson, A., Rauschenbeutel, A. & Volz, J. Beyond the Tavis-Cummings model: Revisiting cavity QED with ensembles of quantum emitters. *Physical Review A* **105**, 013719 (2022). https://doi.org/10.1103/PhysRevA.105.013719
58 Fink, J. M. *et al.* Dressed Collective Qubit States and the Tavis-Cummings Model in Circuit QED. *Physical Review Letters* **103**, 083601 (2009). https://doi.org/10.1103/PhysRevLett.103.083601
59 Amsüss, R. *et al.* Cavity QED with Magnetically Coupled Collective Spin States. *Physical Review Letters* **107**, 060502 (2011). https://doi.org/10.1103/PhysRevLett.107.060502
60 Darman, P. *et al.* Electron beam driven collective self hybridized exciton polaritons. **arXiv:2609.07164** (2026). <https://doi.org/10.48550/arXiv.2609.07164>.
61 Hertzog, M., Wang, M., Mony, J. & Börjesson, K. Strong light–matter interactions: a new direction within chemistry. *Chemical Society Reviews* **48**, 937-961 (2019). https://doi.org/10.1039/c8cs00193f
62 Darbari, S. *et al.* Exciton Energy Routing via Defect Networks in hBN/2D Perovskite Hybrids. *ACS Nano* **20**, 21739-21751 (2026). https://doi.org/10.1021/acsnano.6c03445
63 Abdi, Y. *et al.* 2D Borophene: In-Plane Hyperbolic Polaritons in the Visible Spectral Range. *Advanced Functional Materials* **35**, e13016 (2025). https://doi.org/https://doi.org/10.1002/adfm.202513016
64 Taleb, M., Hentschel, M., Rossnagel, K., Giessen, H. & Talebi, N. Phase-locked photon–electron interaction without a laser. *Nature Physics* **19**, 869-876 (2023). https://doi.org/10.1038/s41567-023-01954-3
65 Taleb, M., Samadi, M. & Talebi, N. Phase-matched electron–photon interactions enabled by 3D-printed helical waveguides. **14**, 3279-3286 (2025). https://doi.org/doi:10.1515/nanoph-2025-0297
66 Chahshouri, F., Taleb, M., Black, M., Mensing, M. & Talebi, N. Cathodoluminescence of polarization-induced energy states at AlGaN/GaN interface. *Journal of Physics D: Applied Physics* **57**, 465104 (2024). https://doi.org/10.1088/1361-6463/ad6fad
67 Black, M. *et al.* Long-Range Self-Hybridized Exciton-Polaritons in Two-Dimensional Ruddlesden–Popper Perovskites. *ACS Photonics* **11**, 4065-4075 (2024). https://doi.org/10.1021/acsphotonics.4c00824
68 Mahr GmbH. *Confocal metrology – 3D surface metrology for industry and research*, 2026).

# Phonon-Enabled Collective Strong Coupling of Color Centers in a Plasmonic Cavity

Bharti Garg[†], Prabhdeep Singh[†], Masoud Taleb, Nahid Talebi*

Institute for Experimental and Applied Physics, Kiel University, Germany

Email: talebi@physik.uni-kiel.de

[†] These authors contributed equally to this work.

## Content:

## Supplementary Note 1: Numerical Simulations

**Modal structure of the cavity –** The void plasmonic cavity manifests (Supplementary Fig. S1) a series of optical modes with both radial (within xy-plane) and vertical (along z-axis) dependencies. The presence of hBN on top of the void cavity, with the background refractive index of $n \simeq 2$, substantially defers the response of the cavity, and as such, photonic modes confined within the cavity become directly excitable by the electron beam. These modes, without the hBN layer on top, do not directly interact with the electron beam, since they are not purely evanescent in nature[1] .

Using our in-house developed finite-difference time domain (FDTD) method [2–6], we first simulated the response of the structure to an electron traversing the cavity in the middle of it, in the absence of emitters. Our electron beam is modelled as a moving Gaussian charge density distribution, with its broadening being only 2 nm both in the transversal and longitudinal

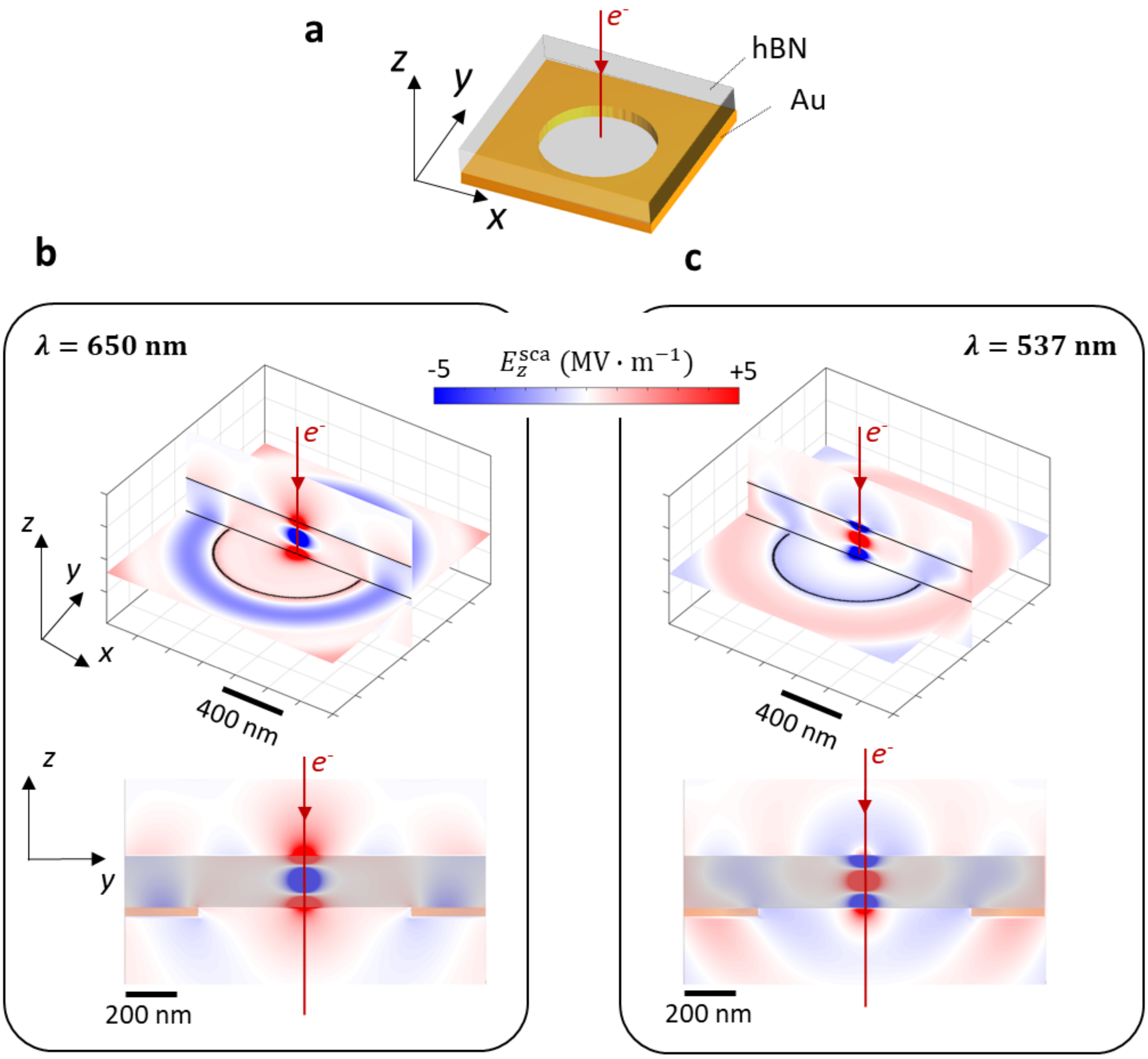

**Supplementary Fig. S1. Modal structure of the void-plasmon cavity. (a)** Structural configuration used in the simulations, encompassing a hole with the diameter of 800 nm inserted in a 40-nm-thick planar gold film, with a 200-nm-thick hBN film inserted on top of the cavity. The z-component of the electric field at **(b)** $\lambda = 650$ nm and **(c)** $\lambda = 537$ nm. Top plots are 3D configurations of the field along the

xy- and xz- planes, whereas the bottom ones are the field configuration in the yz-plane positioned at x=0.

directions with respect to the electron trajectory[7]. For all results discussed below, the simulation domain is fixed as $1400\,\mathrm{nm} \times 1400\,\mathrm{nm} \times 1000\,\mathrm{nm}$, along x, y, and z directions, respectively. For truncating the simulation domain, 2nd order absorbing boundary conditions are implemented that has slightly less efficiency compared to the perfectly-matched layers (PMLs). However, for large simulation domains, this allows us for using the computer capacity predominantly for the structure rather than PMLs. The entire simulation domain is discretised with 1 nm unit cells along all 3 axes. For modelling gold, the experimental permittivity is fitted to a Drude model + two critical point functions[8–10] . The initial electron kinetic energy is 30 keV. The Maxwell simulator is combined with the Lorentz equations, in order to account for possible electron recoils and modifications to the trajectories due to the interactions with defects and/or the electron-induced scattered field[11] .

The scattered z-component of the electric field at the resonant wavelengths of 537 nm and 650 nm is shown in Supplementary Fig. S1. The scattered field is clearly enhanced along the electron trajectory. Due to the low refractive index of hBN, no Cherenkov radiation is generated. The main optical cavity mode that resonantly interacts with emitter 2, as well as with the fourth vibronic mode of emitter 1, corresponds to the resonance at 537 nm (see Fig. 3b). The optical mode at 650 nm exhibits a nearly uniform field distribution within the hole, whereas the mode at 537 nm shows a higher-order radial field dependence. Both resonances exhibit a second-order mode profile along the z-axis. In contrast, the longer-wavelength resonance at 960 nm exhibits a first-order mode profile both along the *z*-axis and in the radial direction.

**Maxwell-Bloch Simulations –** For calculating the polarization induced by the emitters in the structure, the FDTD simulator for the Maxwell equations is merged with a solver for numerically simulating the dynamics of the density matrix ( $\hat{\rho}$ ) via the master equation as below:

$$\frac{d\hat{\rho}_j}{dt} = -\frac{i}{\hbar}\left[\hat{H}_j, \hat{\rho}_j\right] + \hat{L}\left\{\hat{\rho}_j\right\}, \tag{1}$$

where $\hbar$ is the reduced Planck Constant, $\hat{H}_j$ is the system Hamiltonian associated with the $j^{\text{th}}$ emitter, expressed as

$$\hat{H}_j = \sum_{n=1}^{N_j} \hbar\omega_n^j \left|n\right\rangle_j \left\langle n\right|_j - \vec{\hat{\mu}}^j \cdot \vec{E}\left(\vec{r}_j, t\right), \tag{2}$$

and $\hat{L}\left\{\hat{\rho}\right\}$ is the Lindblad operator, expressed as

$$\hat{L}\left\{\hat{\rho}_j\right\} = \sum_{m<n} \gamma_{mn}^j \left(\hat{\sigma}_{mn}^{j-}\,\hat{\rho}\,\hat{\sigma}_{mn}^{j+} - \frac{1}{2}\left\{\hat{\sigma}_{mn}^{j+}\hat{\sigma}_{mn}^{j-}, \hat{\rho}\right\}\right), \tag{3}$$

with $\gamma_{mn}^j$ being the radiative decay associated with transitions from states $\left|m\right\rangle_j$ to $\left|n\right\rangle_j$ of the $j^{\text{th}}$ emitter. $\hat{\sigma}_{mn}^{j-} = \left|m\right\rangle_j \left\langle n\right|_j$ and $\hat{\sigma}_{mn}^{j+} = \left|n\right\rangle_j \left\langle m\right|_j$ are the lowering and raising operators for each emitter. Furthermore, in equation (2), $\hbar\omega_n^j$ is the transition energy from state $n^{\text{th}}$ to the ground state and $\vec{\hat{\mu}}^j$ is the dipole matrix operator, for each emitter *j*. $\vec{E}\left(\vec{r}_j, t\right)$ is the total electric field at the position of the $j^{th}$ emitter, including the field generated from all the emitters within the ensemble, except the field generated from the $j^{th}$ emitter itself.

Hence, the interactions between different emitters are only included as photon-mediated energy exchange, which depending on the relative distance between the emitters and their positioning, can be dipolar near-field effect or far-field interactions. In close analogy with the experiments, we consider two sets of emitters.

First, a number of 200 of emitter type 1 which is modeled by a 5-level system (Supplementary Fig. 2b), are distributed randomly near to each surface of the hBN at only up to a 10 nm distance from the top and bottom surfaces, since our depth-resolved cathodoluminescence (CL) spectroscopy results demonstrate that they are positioned near to the surface. Because these emitters are generated during the liquid exfoliation process, we insert these emitters in our simulations both near to the top and bottom surface of the hBN (red closed circles in Supplementary Fig. 2a). The transition wavelengths for emitter 1 is fixed as $\lambda_1 = 440\,\mathrm{nm}$, $\lambda_2 = 460\,\mathrm{nm}$, $\lambda_3 = 500\,\mathrm{nm}$, and $\lambda_4 = 537\,\mathrm{nm}$. The radiative damping between states are set to $\gamma_{01} = 8\cdot 10^{13}\,\mathrm{Hz}$, $\gamma_{03} = 5\cdot 10^{13}\,\mathrm{Hz}$, $\gamma_{04} = 4\cdot 10^{13}\,\mathrm{Hz}$, and the rest of the elements are considered to be $2\cdot 10^{13}\,\mathrm{Hz}$. The dipole matrix elements for this emitter are set to $\mu_{01} = 10\,\mathrm{D}$, $\mu_{02} = \mu_{03} = \mu_{04} = 8\,\mathrm{D}$, $\mu_{12} = \mu_{13} = \mu_{14} = 5\,\mathrm{D}$, and $\mu_{13} = \mu_{24} = \mu_{34} = 2\,\mathrm{D}$.

Second, we consider 500 emitters of type 2, modelled by a 3-level system, and randomly position them within the bulk of the hBN. The positioning procedure is chosen in such a way that the emitters should be positioned at an average inter-distance of 100 nm with respect to each other and with a variance of 30 nm, to avoid direct strong correlations between the emitters, via near-field interactions. The transition wavelengths, radiative damping, and the dipole matrix elements associated with this emitter is set as $\lambda_1' = 537\,\mathrm{nm}$, $\lambda_2' = 600\,\mathrm{nm}$,

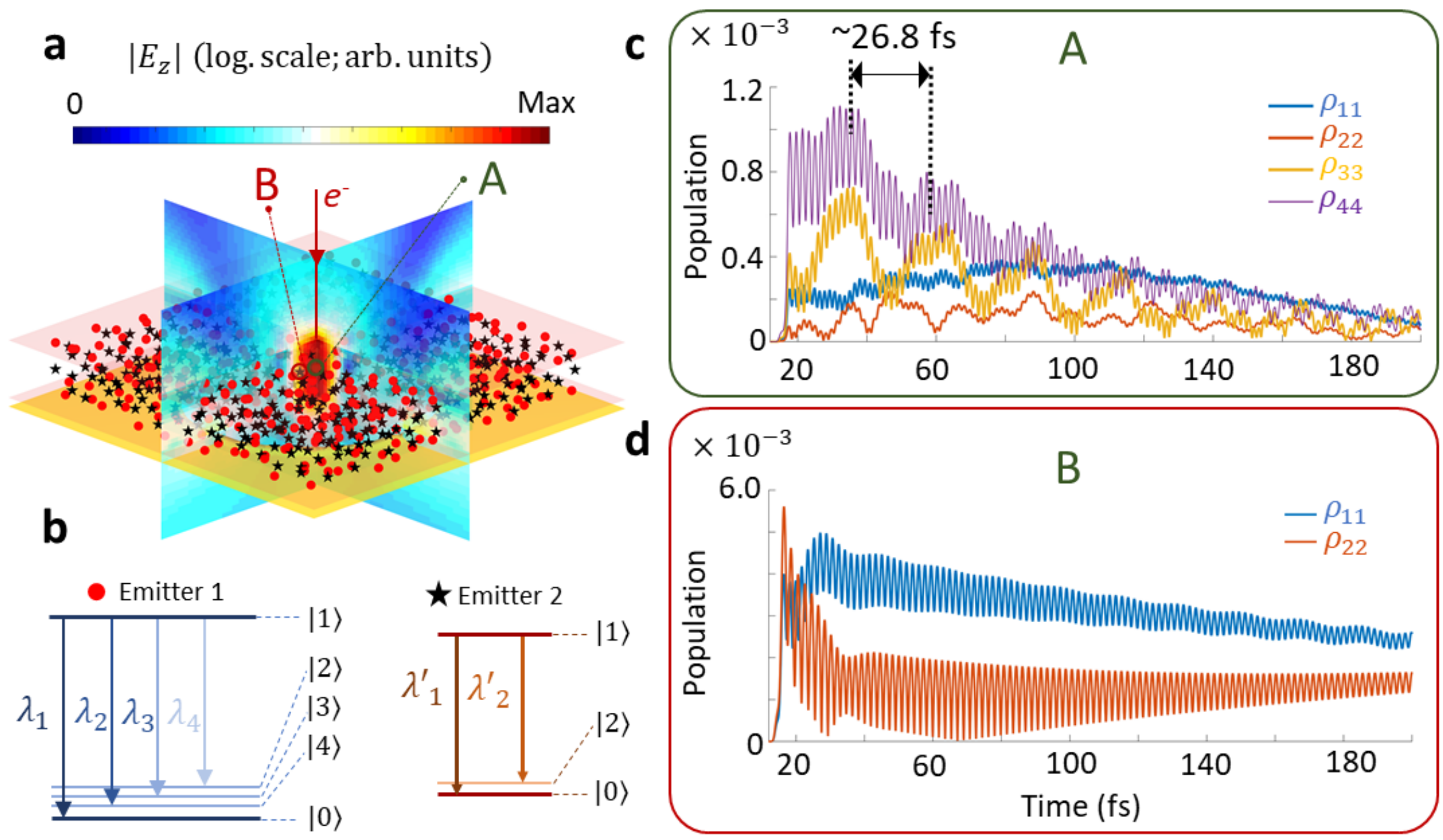

**Supplementary Fig. S2. Maxwell-Bloch Simulation Framework. (a)** An ensemble of emitters are positioned at the hBN/vacuum and hBN/gold interfaces within 20 nm in depth, as well as within the bulk

of the hBN, for emitter types 1 and 2 respectively. Displayed colored map is the z-component of the electric field at the wavelength of 537 nm, presented in the logarithmic scales. The hBN/vacuum and hBN/gold interfaces is shown by the pink planes, whereas the bottom layer of the gold film is shown by the yellow plane. (b) Energy diagram of type 1 and type 2 emitters implanted in the simulations. For concrete parameters of the emitters see the text. Population dynamics of the higher energy states for (c) type 1 and (d) type 2 emitters, for the emitters marked as A and B in (a).

$\gamma'_{01} = \gamma'_{12} = 6 \cdot 10^{13}\,\mathrm{Hz}$, $\gamma'_{02} = 8 \cdot 10^{13}\,\mathrm{Hz}$, and $\mu'_{01} = 10\,\mathrm{D}$, $\mu'_{02} = 5\,\mathrm{D}$, and $\mu'_{12} = 2\,\mathrm{D}$ (in Debye units).

Equation (1) to (3) are solved numerically using a 4th order Runge-Kutta scheme, implemented in each time step of the FDTD method. The calculated polarization

$$P(\omega) = \frac{e^2}{3\varepsilon_0 \hbar} \sum_{j=1}^{N_{\mathrm{em}}} \int_{-\infty}^{+\infty} dt e^{-i\omega t} tr\left\{\hat{\mu}_j \hat{\rho}_j(t)\right\}, \quad (4)$$

where $N_{\mathrm{em}}$ is the total number of emitters, is self-consistently added to the polarization field within the FDTD scheme, to account for photon-mediated interactions within the emitters. More precisely, within each time step, we first calculate the force inserted on the electron beam (Lorentz force), the electron position, its current density distribution, and successively update the magnetic field, add the incident field when considering optical excitation schemes, the displacement vector, the polarization, the electric field, the master equation (2) using the Runge-Kutta scheme, add the emitter polarization to the electron- (or light-) induced polarization, and finally apply the absorbing boundary conditions.

The results for a moving electron at the kinetic energy of 30 keV is demonstrated in Figure 4. The time-dependence population dynamics for two emitters belonging to the first and second types are plotted in the Supplementary Fig. S2 c and d, respectively. The populations sharply rise from 0 to large values at 12 fs, when the electron starts traversing the structure. Even if the interaction time of the electron with the entire structure is only 2.43 fs, the populations decay takes place at time scales as long as 100 fs for emitter type and 180 fs for emitter type 2. For type 1 emitter, one notices that the build-up of populations takes place with first occupying the phonon lines, and sequentially the ZPL, whereas for emitter type 2, the situation is vice versa. This could be explained by the fact that the resonance frequency of the plasmonic structures is tuned with the 4th vibrational state of emitter type 1 and the ZPL of emitter type 2. For emitter type 1, strong modulations in $\rho_{44}(t)$ as well as $\rho_{33}(t)$ (purple and yellow lines in Supplementary Fig. S2 c), at the order of 26.87 fs, substantiates the observed energy split in the spectra at the order of $\Delta E = \hbar(\Delta t)^{-1} \simeq 24.5\,\mathrm{meV}$, which is in close agreement with the experimentally observed energy split in the CL spectrum. Interestingly, the modulations observed in the population dynamics associated with the 2nd emitter type are less pronounced, even if the ZPL transition wavelength is perfectly matching the resonance wavelength of the plasmonic structure.

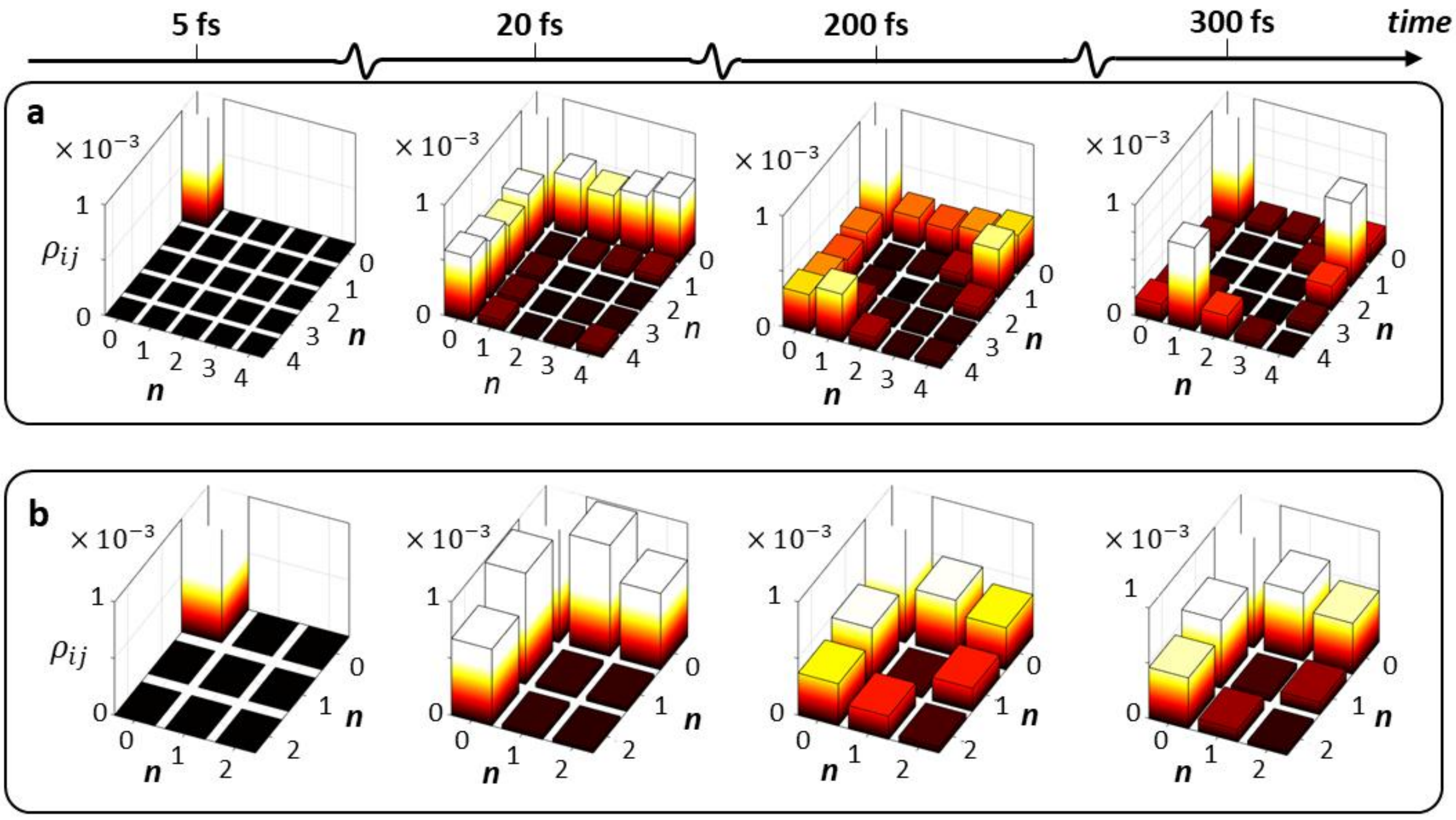

**Supplementary Fig. S3. Dynamical evolution of the density matrices.** Illustrated matrix elements at depicted times for **(a)** emitter 1, and **(b)** emitter 2, marked as A and B in Supplementary Fig. S2 a, respectively.

The coherence terms in the density matrix, that substantiates the emitter-induced polarization (see Eq. (4)), better demonstrates the reason for observing the energy splitting behaviour for the vibrational modes (See Supplementary Fig. S3). Indeed, the most prominent and long-lasting coherence term in the density matrix is the $\rho_{14}$ term associated with the emitter type 1.

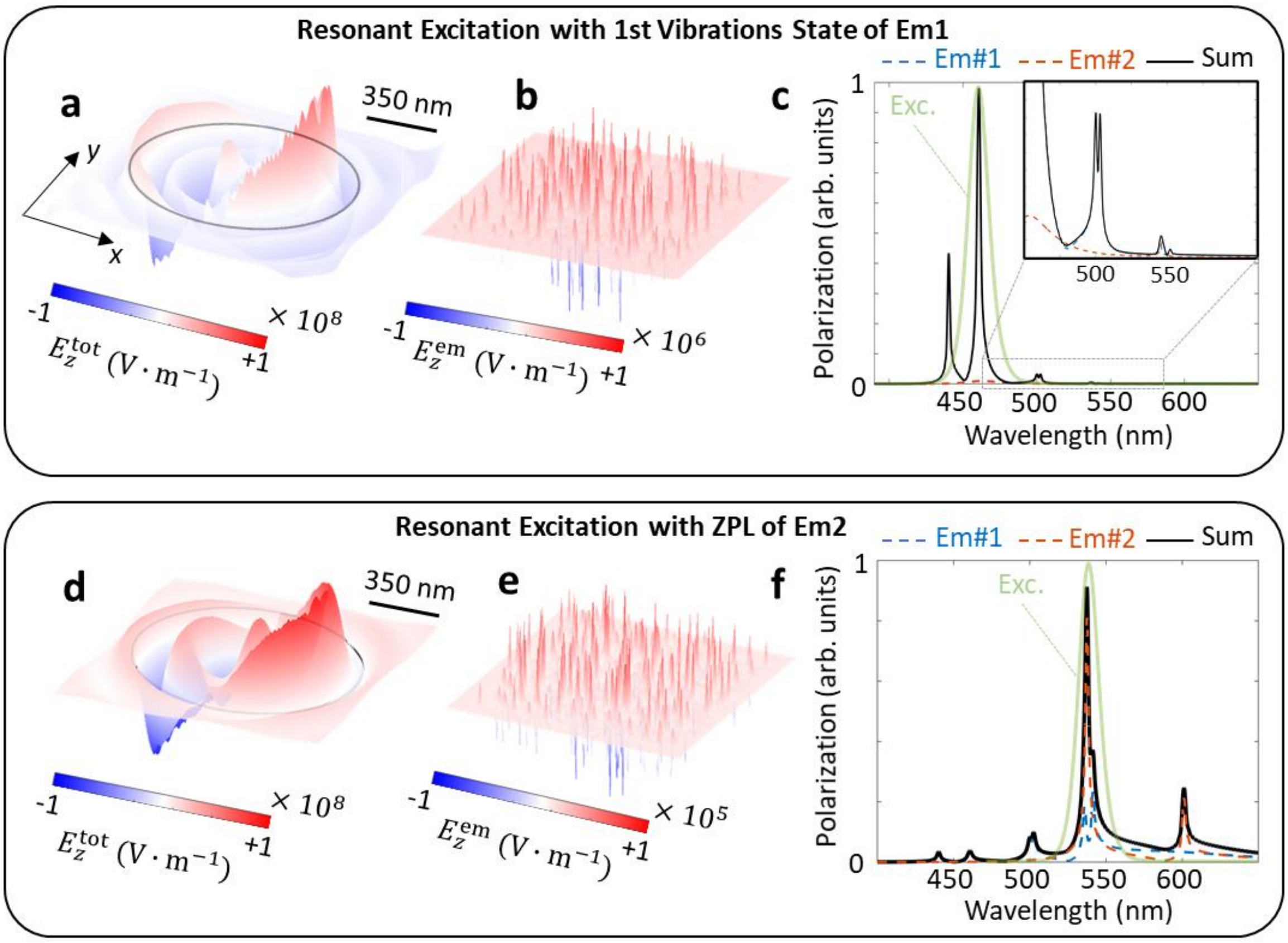

**Supplementary Fig. S4. Excitation with light.** z-component of the electric field at 5 nm above the cavity within the hBN layer, for **(a)** total field, and **(b)** induced field by the emitters, for the excitation wavelength of 460 nm, **(d)** total field, and **(e)** induced field by the emitters, for the excitation wavelength of 537 nm. Calculated polarization spectra for **(c)** excitation wavelength of 460 nm and **(f)** excitation wavelength of 537 nm. The total induced polarization by emitters 1 and 2 are depicted by dashed blue and dashed orange lines, respectively. The summed polarizations are demonstrated by solid black lines, and the excitation field intensity is shown by the solid green line.

Furthermore, excitation with light as compared to electron beams, as well as leads to the emergence of the energy split in the polarization spectrum (Supplementary Fig. S4). Similar to electron-beam excitation, both 3$^{rd}$ and 4$^{th}$ vibrational modes demonstrate a split in their energies. First, we consider an x-polarized spatially and temporally Gaussian light pulse with the full-wave-half-maximum (FWHM) temporal broadening of 7.7 fs, beam waist of 1.2 µm, and the center wavelength of 460 nm interacting with the system. The excitation field is therefore tuned with the 1$^{st}$ vibration state of emitter type 1, similar to the performed photoluminescence measurements (see Figure 3). Even if the excitation wavelength is not tuned with the resonant energy of the plasmonic cavity, the energy split is fully retrieved. This phenomenon hints for the fact that the emitter's self-field interacts with the cavity: In this excitation configuration, the excited field first initiate the transition from the 2$^{nd}$ vibration state to the highest energy state, consequently lead to sequential transitions and emissions of photons via transitions from the highest energy state to the 3$^{rd}$ and 4$^{th}$ vibration states. The emitted photons interact with the plasmonic cavity and lead to the transfer of energy between the plasmonic field and the emitters. Interestingly, the observed energy split does not alter by changing the excitation scheme; e.g., by changing the excitation center wavelength from 460 nm to 537 nm, we again see a wavelength split of 5 nm, corresponding to an energy split of 21.5 meV, which is in a good agreement with the experimentally observed value of 24.09 meV (Supplementary Fig. S4).

The strong coupling factor between a single emitter and a cavity is described as follows. The entire Hamiltonian of the system within the minimal-coupling framework is derived as

$$\hat{H} = \hbar\omega_c \hat{a}^\dagger \hat{a} + \hat{H}_{\mathrm{em}} + \frac{e}{m_0}\hat{\mathbf{A}}\cdot\hat{\mathbf{p}}, \tag{5}$$

where the first term is associated with the field in the cavity, the second term with the emitter, and the third term is the coupling between the cavity and the emitter, where for deriving the latter, the Coulomb gauge is considered. $\hat{a}$ and $\hat{a}^\dagger$ are the photon annihilation and creation operators. The dipole matrix associated with the emitter transitions is given by $\hat{\vec{\boldsymbol{\mu}}} = e\hat{\mathbf{r}} = \mu_{eg}\left(|e\rangle\langle g| + |g\rangle\langle e|\right)\vec{\boldsymbol{\mu}}$, where $|g\rangle$ and $|e\rangle$ are the ground and excited sates of the emitter, respectively. The momentum operator is recast in the form:

$$\begin{aligned}\hat{\mathbf{p}} &= m_0\dot{\hat{\mathbf{r}}} = m_0\frac{1}{i\hbar e}\left[e\hat{\mathbf{r}}, \hat{H}_a\right] = \frac{m_0}{i\hbar e}\left[\hat{\vec{\boldsymbol{\mu}}}, \hat{H}_a\right] = \\ &= \frac{m_0}{i\hbar e}\frac{1}{2}\hbar\omega_a\mu_{eg}\vec{\boldsymbol{\mu}}\Big(\left(|e\rangle\langle g| + |g\rangle\langle e|\right)\left(|e\rangle\langle e| - |g\rangle\langle g|\right) \\ &\qquad - \left(|e\rangle\langle e| - |g\rangle\langle g|\right)\left(|e\rangle\langle g| + |g\rangle\langle e|\right)\Big) \\ &= \frac{m_0}{i\hbar e}\hbar\omega_a\mu_{eg}\vec{\boldsymbol{\mu}}\left(|g\rangle\langle e| - |e\rangle\langle g|\right) = i\frac{m_0}{\hbar e}\hbar\omega_a\mu_{eg}\vec{\boldsymbol{\mu}}\left(\hat{b}^\dagger - \hat{b}\right),\end{aligned} \tag{6}$$

where $\hat{b} = |e\rangle\langle g|$ and $\hat{b}^\dagger = |g\rangle\langle e|$, respectively. The vector potential operator is given by

$$\hat{\mathbf{A}} = \hat{\mathbf{e}}\sqrt{\frac{\hbar}{2\varepsilon_0\omega_c V}}\left(\hat{a}^\dagger + \hat{a}\right). \tag{7}$$

Here, $V$ is the mode volume of the cavity, $\varepsilon_0$ is the free-space permittivity, and $\hat{\mathbf{e}}$ is the field polarization. Inserting equations (6) and (7) in eq. (5) result in

$$\hat{H} = \hbar\omega_c\hat{a}^\dagger\hat{a} + \hbar\omega_e\hat{b}^\dagger\hat{b} + i\hbar g_1\left(\hat{a}^\dagger + \hat{a}\right)\left(\hat{b}^\dagger - \hat{b}\right), \tag{8}$$

with the strong coupling factor

$$E_{g,1} = \hbar g_1 = \left(\vec{\mathbf{e}}\cdot\vec{\boldsymbol{\mu}}\right)\sqrt{\frac{\hbar\omega_a}{2\varepsilon_0 V}}\mu_{eg}\ . \tag{9}$$

Considering that the mode volume is approximately given by $V = \pi r^2 h$, where $r = 400\,\mathrm{nm}$ is the radius of the plasmonic cavity and $h = 10\,\mathrm{nm}$ is the penetration depth of the plasmonic field in the hBN layer, the coupling strength is given by only approximately $E_{g,1} = 0.0014\ \mathrm{eV}$, which corresponds to an energy split of only $2E_{g,1} = 2.8\ \mathrm{meV}$ under resonant excitation conditions. However, for a collective excitation of all 200 type-2 emitters, the coupling strength is modified by $E_{g,N} = \sqrt{N}\hbar g_1 = 20$ meV, corresponding to a resolved energy split of 40 meV. The numerically-resolved split is 24.5 meV, since not all the emitters experience a synchronized and collective dynamics. In fact, we observe that those emitters that are positioned nearer to the rim of the structure experience a strongly synchronized modulation of their population. This number is resolved as $N = (12.25\,\mathrm{meV}/\hbar g_1)^2 \approx 80$ emitters.

## Supplementary Note 2: Stokes Shift

PL measurements are performed at the same positions investigated by CL, directly on the hBN flake above the Au plasmonic void cavity, to determine whether the cavity-induced hybridized resonance can also be accessed under optical excitation; thus, allowing the role of phonon-assisted excitation processes to be examined. The accessible excitation range was limited by the dichroic mirror and long-pass filter combination (cut-off ≈490 nm), preventing the use of longer excitation wavelengths closer to the 537 nm emission. Only positions where all Stokes shifts could be reliably extracted are plotted.

Supplementary Fig. S5 summarizes the PL spectra from the excitation wavelengths between 465 and 485 nm together with the extracted Stokes shifts. The Au void cavity supports localized plasmonic resonances that enhance the Raman scattering efficiency, enabling the observation of well-resolved Stokes-shifted emission[12]. Two dominant phonon-assisted energy shifts of approximately 0.17 and 0.20 eV are observed. These energies are consistent with activation of the IR-active TO and LO phonon branches of hBN[13], as explained in the main text. The slightly larger deviation of the lower-energy shift could be attributed to the partially unresolved first-order Stokes feature, whereas the higher-energy component agrees closely with the expected LO phonon energy.

In addition to the individual phonon-assisted transitions, the spectra exhibit a combined Stokes shift of approximately 0.37 eV, corresponding to the simultaneous participation of both phonon modes. At an excitation wavelength of 465 nm (2.66 eV), the combined energy loss of approximately 0.37 eV results in a down converted emission at the energy of 2.29 eV (≈540 nm), which coincides with the vibronic transition participating in the cavity coupling. The phonon-assisted down-conversion therefore provides an efficient excitation pathway.

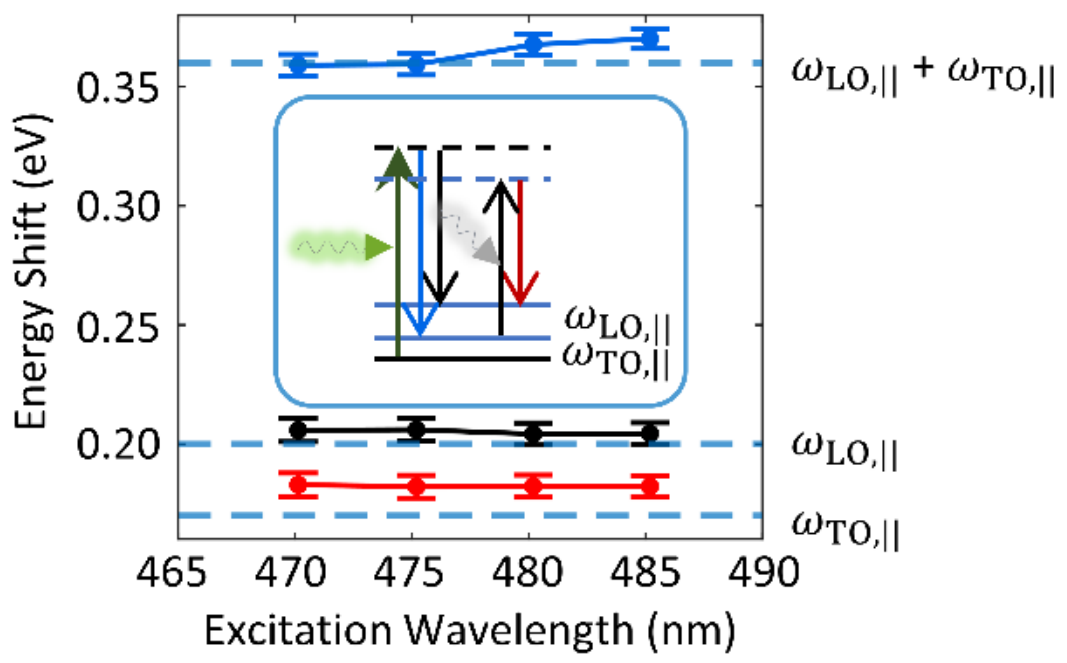

**Supplementary Fig. S5. Stokes shifts energies extracted from the PL spectra under excitation wavelengths from 470 to 485 nm.** The dashed lines indicate the TO and LO phonon energies and their combined energy. The inset illustrates the phonon-assisted relaxation pathways.

## Supplementary Note 3: CL and PL Energy Splits

To directly compare the cavity-induced hybridized resonance under electron- and photon-excitation, the split spectral feature observed in CL and PL is shown in Supplementary Fig. S6.

The two measurements were performed at the same cavity position, allowing a direct comparison of the hybridized state under identical sample conditions.

The CL spectrum exhibits two maxima centered at 538.74 nm and 547.36 nm, corresponding to an energy splitting of 36.24 meV. Under PL excitation, the split resonance is observed at 538.53 nm and 544.42 nm, giving a reduced splitting of 24.09 meV. While the higher-energy component remains nearly unchanged in both measurements, the lower-energy branch shifts towards shorter wavelengths in PL, resulting in the smaller energy difference.

The PL spectrum shows a more clearly resolved doublet than the CL spectrum, that could be explained by the selective photon energy excitation in PL compared to the broadband carrier generation in CL. In contrast, CL simultaneously excites multiple electronic transitions and generates a larger incoherent background, leading to broader spectral features and partial overlap of the hybridized branches. Despite this, both excitation schemes consistently reveal the same cavity-induced hybridized resonance.

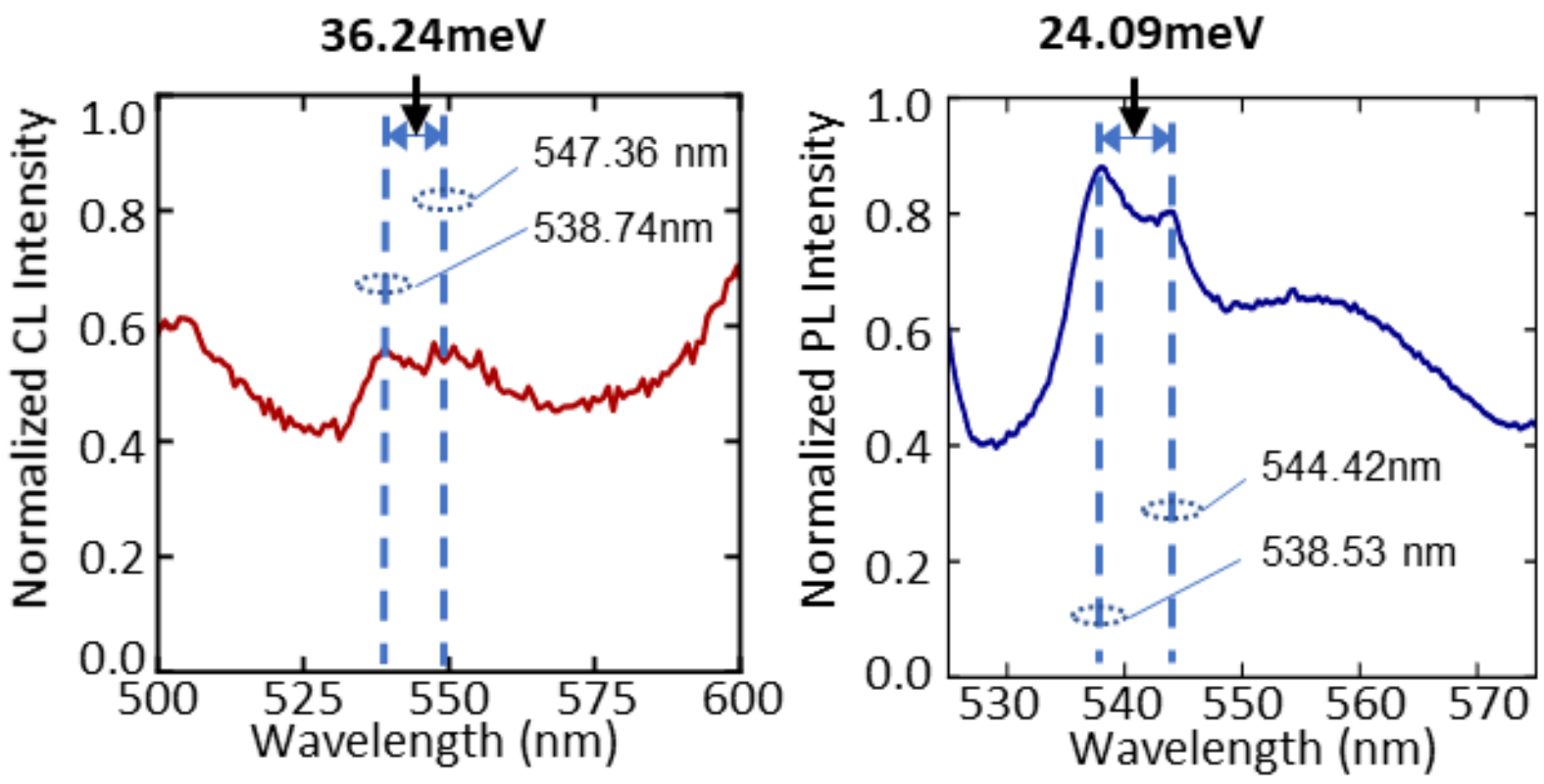

**Supplementary Fig. S6. Normalized CL and PL spectra measured at the same cavity position, showing the cavity-induced spectral splitting.**

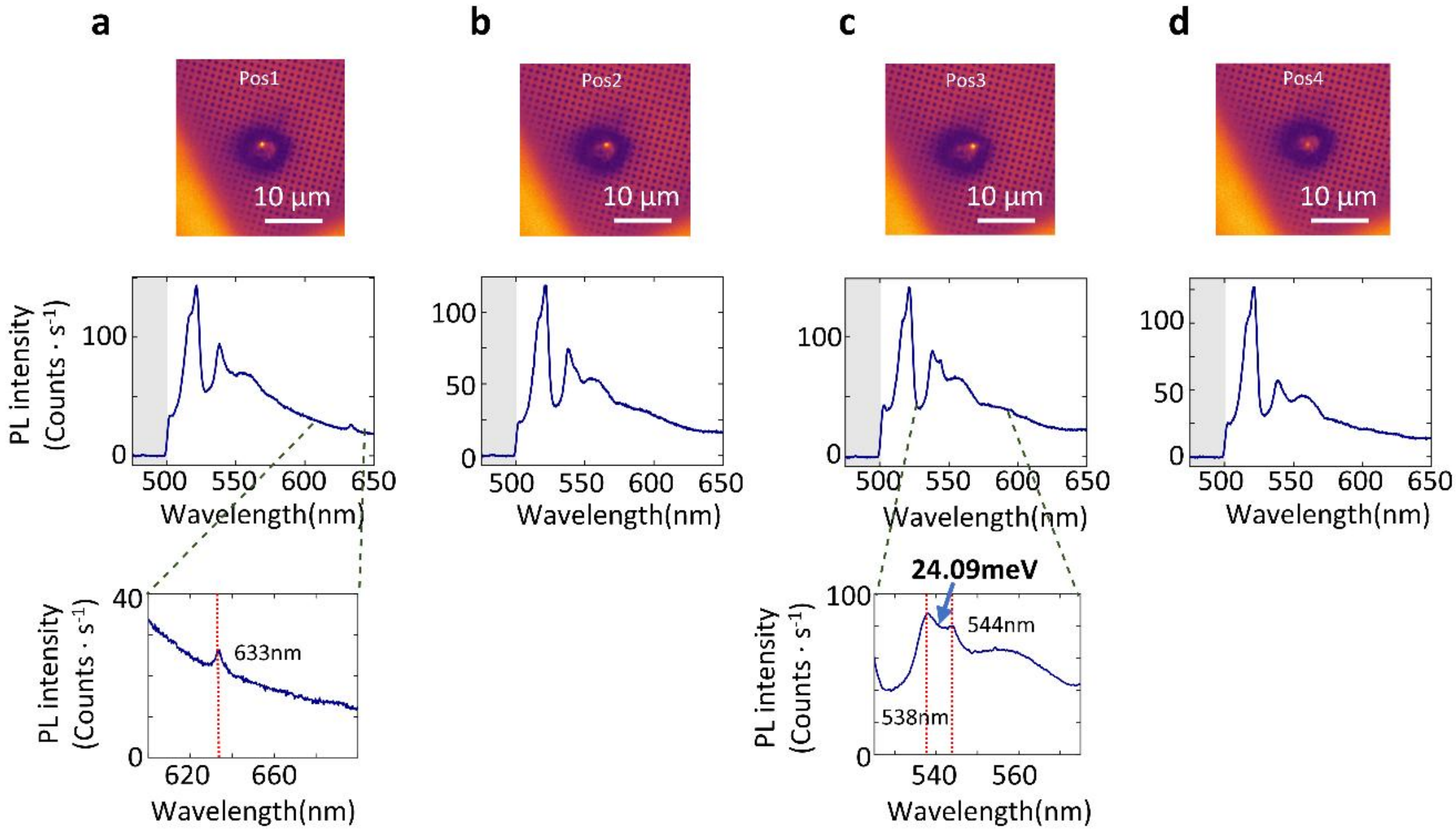

**Supplementary Fig. S7.** (a-d) Optical microscope images of the investigated hBN flake showing the excitation laser spot at measurement positions Pos1-Pos4 with corresponding PL spectra taken at each position.

## Supplementary Note 4: Position-Dependence of the Cavity-Emitter Interactions

PL measurements are performed at several locations across the hBN flake. Among the investigated positions, only a small number located directly above the Au void cavity exhibited the characteristic hybridized spectral response (Supplementary Fig. S7). The hybridized spectral doublet, with peaks at 538.53 and 544.42 nm separated by 24.09 meV, is observed exclusively when the laser is focused on the cavity (Pos3). In contrast, spectra recorded only a few micrometers away (Pos1, Pos2, and Pos4) exhibit neither the mode splitting nor the narrow cavity-coupled resonance, despite the similar excitation conditions. Besides the 538 nm emission, Pos1 also show an emission feature near 633 nm.

In contrast, the 538 nm defect emission and its associated Stokes-shifted features remain observable over much of the flake, indicating that these defect centers are more broadly distributed. Nevertheless, the cavity-induced hybridized resonance is detected only at the cavity position, showing that spectral overlap alone is not sufficient to produce the coupled state and a strong spatial overlap with the appropriate emitter density is required.

## Supplementary Note 5: Excitation Efficiency and Flake Morphology

To correlate hBN morphology with the emission spectra, thickness maps measured using a MarSurf confocal microscope were combined with PL and CL measurements performed at the same positions across several representative hBN clusters.

Supplementary Fig. S8 presents an hBN cluster where three measurement positions were selected along the black arrowheads marked in the thickness map. Two positions (Pos2 and Pos3) are located in regions thicker than 1 µm, whereas the third position (Pos1) is approximately 300 nm thick. Pos1 shows the weakest background together with relatively strong defect emission. In contrast, Pos2 and Pos3 both show a strong broadband background with only weakly resolved defect emission.

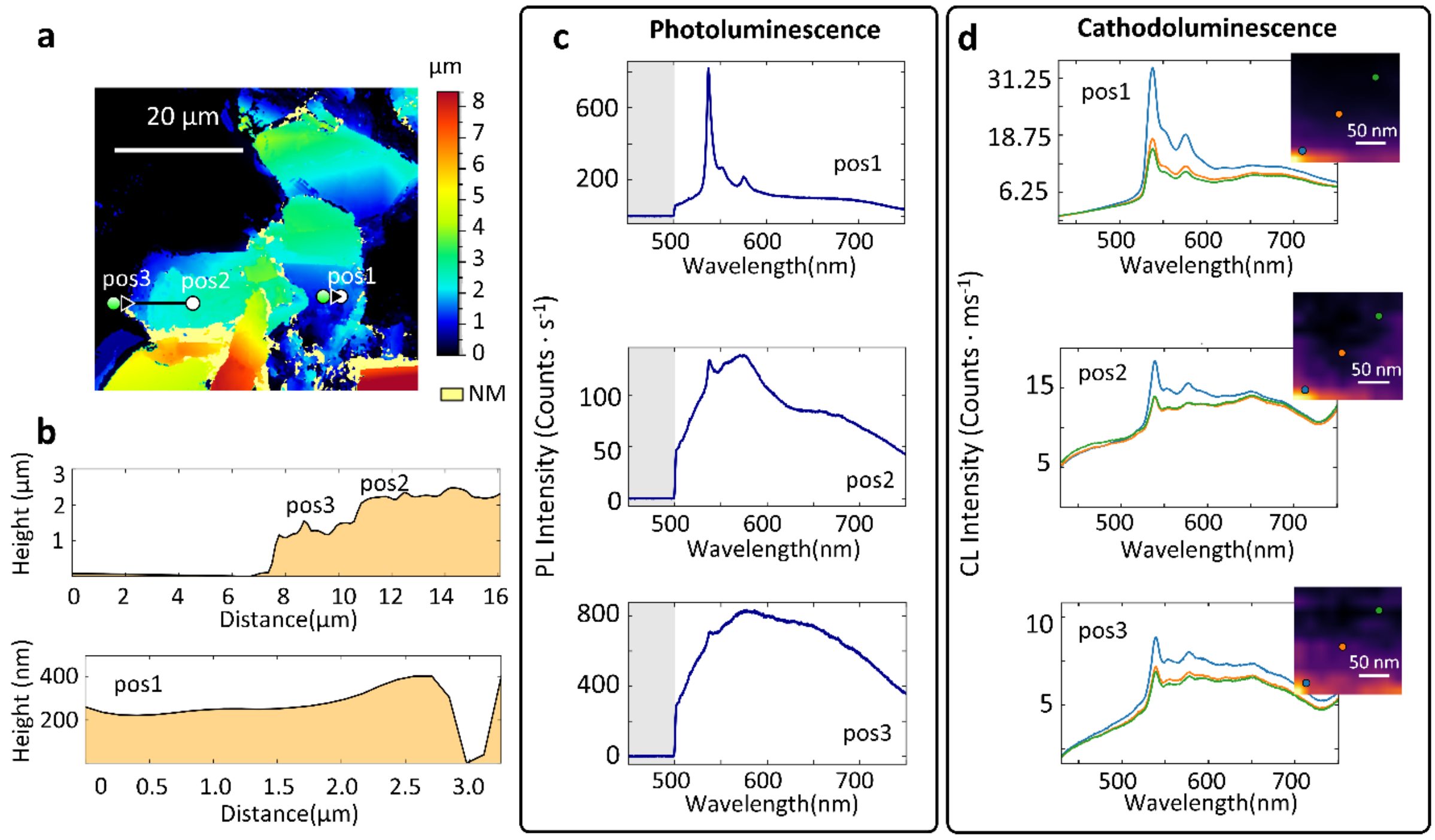

**Supplementary Fig. S8. Thickness-dependent PL and CL measurements on a representative hBN cluster.** (a) Confocal thickness map showing the measurement positions. (b) Height profiles along the indicated lines. (c) PL spectra and (d) CL spectra measured at the corresponding positions.

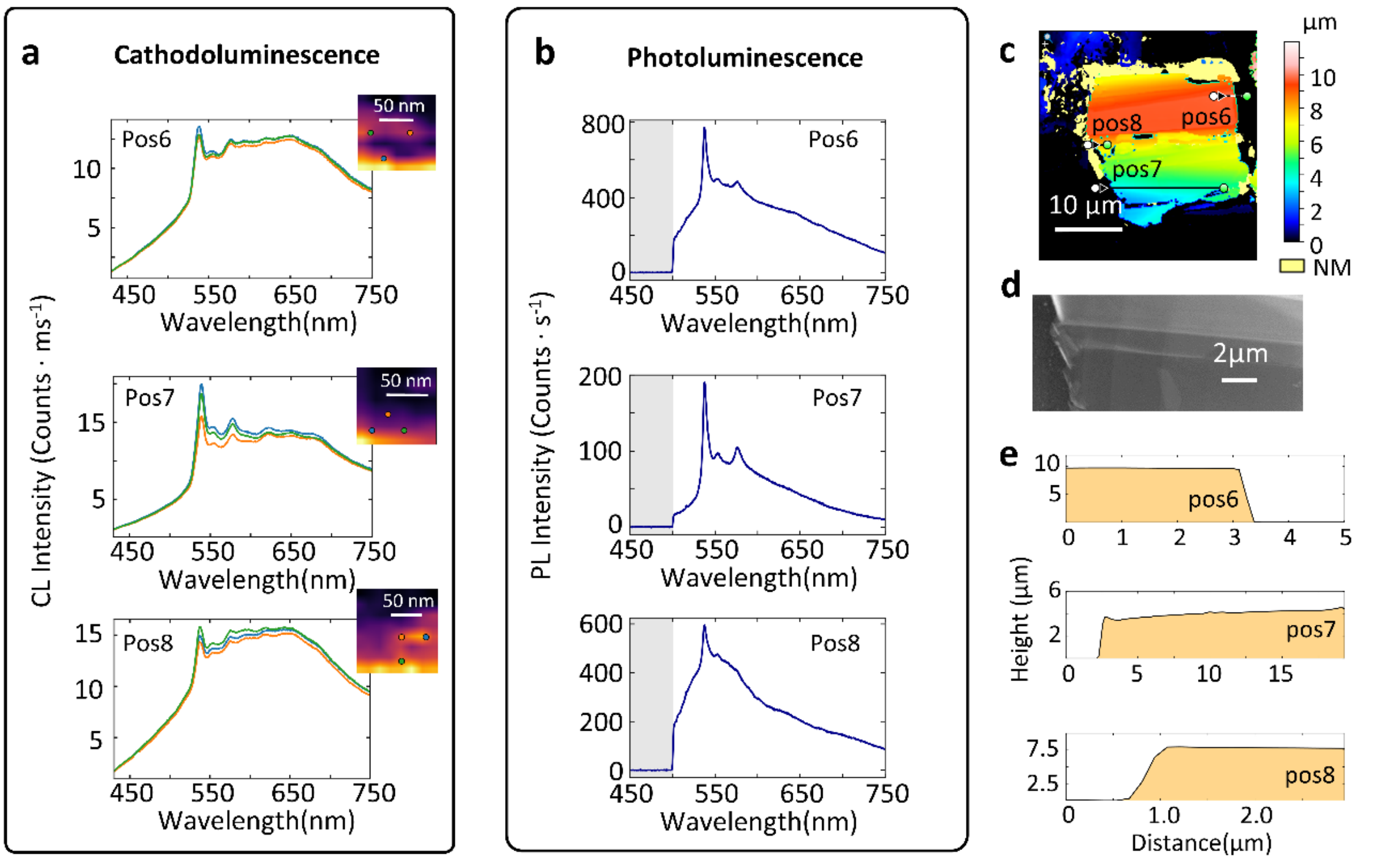

**Supplementary Fig. S9. PL and CL spectra measured at representative positions with different surface morphology and flake thickness**. The corresponding (c) thickness map, (d) SEM image, and (e) height profiles are shown on the right.

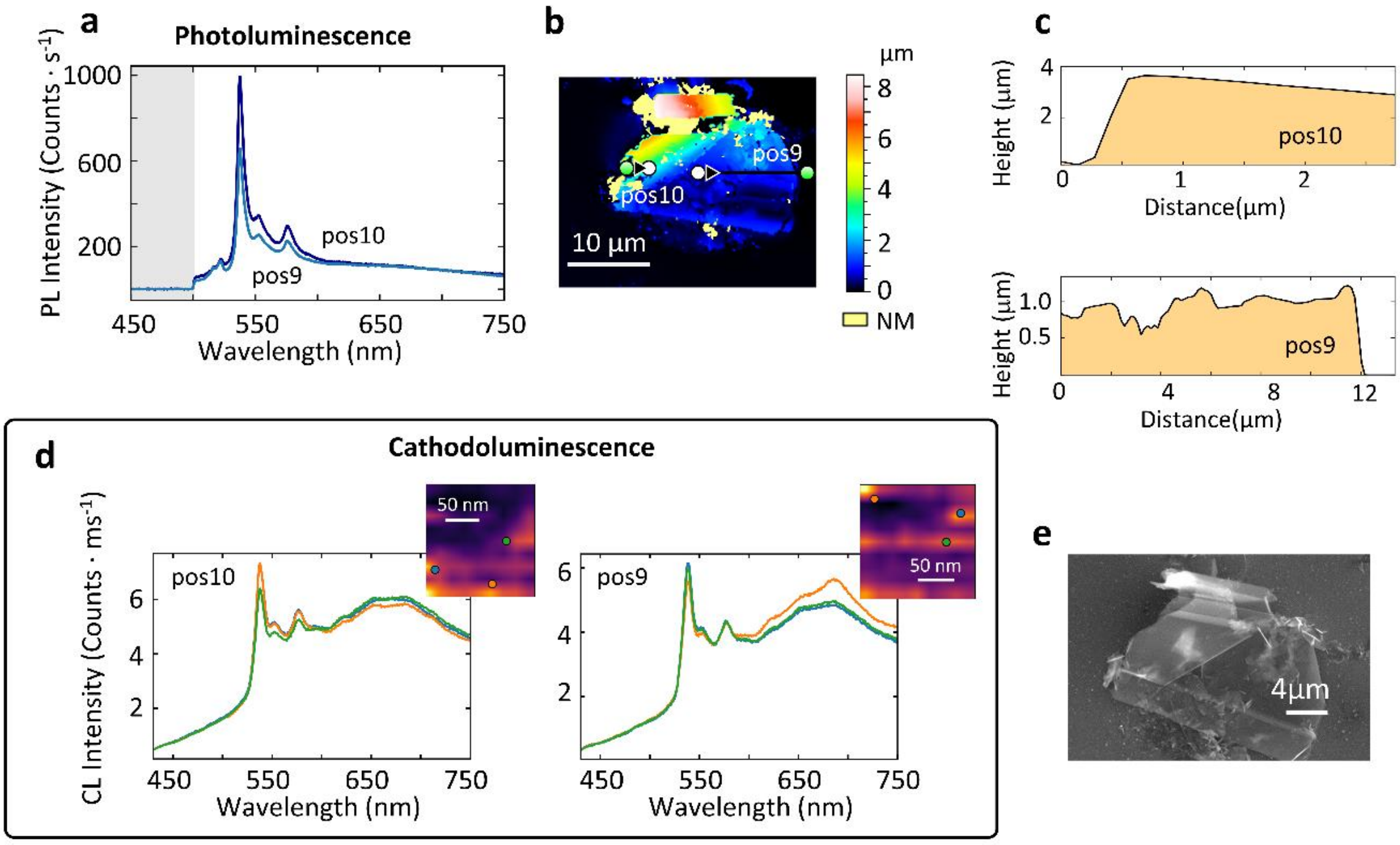

**Supplementary Fig. S10**. **Effect of Flake Morphology on the collection efficiency**. (a) PL and (d) CL spectra measured at different positions on rough hBN flake. The corresponding (b) thickness map, (e) SEM image, and (c)height profiles are shown.

Additionally, hyperspectral CL maps integrated over the 530-540 nm wavelength range were taken to visualize the spatial distribution of the defect emission. The corresponding hyperspectral maps are presented in the inset of each CL spectra. The emission is governed by both the local morphology and the flake thickness[14]. Sample morphology influences the emission efficiency by controlling the random scattering and outcoupling of light emitted from the color centers, while increasing flake thickness enhances the incoherent phonon-induced broadband background.

Together, these effects determine the relative visibility and outcoupling efficiency of the defect emission in both PL and CL. The same behavior is reproduced for additional hBN clusters (Figs. S9-S11). Supplementary Fig. S9 (Pos6, Pos7, and Pos8) shows measurements on thicker flakes with thicknesses ranging from 4-10 µm, where both PL and CL show a stronger background with increasing thickness. Supplementary Fig.S10 (Pos9 and Pos10) shows measurements from a different hBN cluster. Although Pos9 has a thickness of only 1 µm, its relatively rough surface results in a background signal comparable to that of the 4 µm-thick region (Pos10). This indicates that surface roughness also increases the background by random scattering and producing a less localized excitation volume. Supplementary Figure S11 (Pos4 and Pos5) shows hBN flake thicker than 10 µm, where the defect signal is barely resolved and the spectra are dominated by the background. PL consistently resolves the defect emission slightly better than CL due to the wavelength selective optical excitation. Overall, these measurements helped us identify flakes with lower background emission and prioritize regions with stronger defect emission.

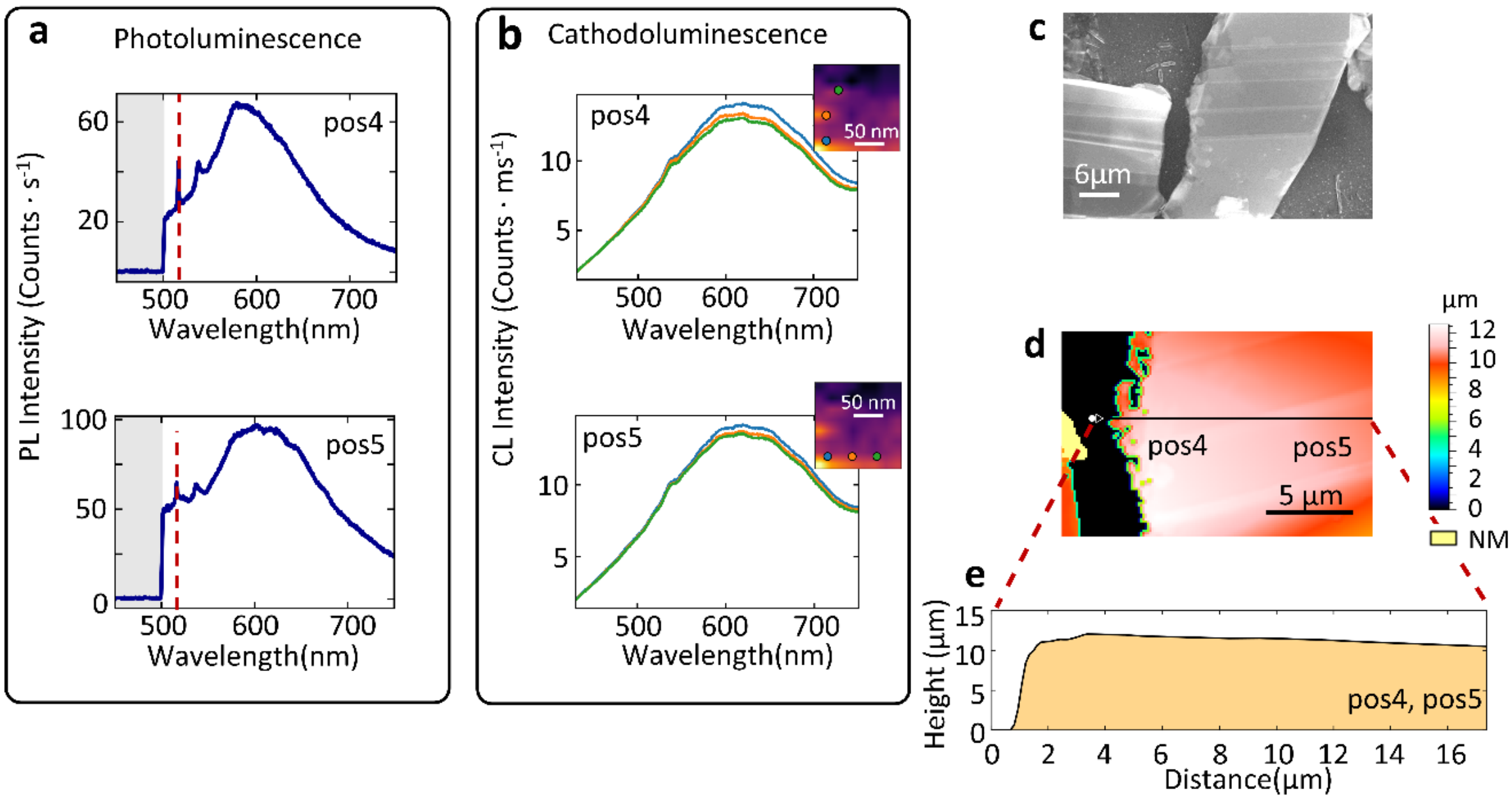

**Supplementary Fig. S11. Stokes shift and emitters in ultrathick hBN flakes. (a,b)** PL and CL spectra measured at different positions on thick hBN flake. The corresponding **(c)** SEM image, **(d)** thickness map and **(e)** height profiles are shown.

## Supplementary Note 6: Distance-dependent CL Measurements

Supplementary Fig. S12a shows the SEM image of an hBN flake together with hyperspectral CL map extending up to 30 µm away from the flake. The CL spectra measured at different beam positions are shown in Supplementary Fig. S12b, where the corresponding measurement positions are marked on the hyperspectral CL image. The defect emission centered around 537 nm remains observable over the entire scan range, showing that the hBN defects can be excited even when the primary electron beam is positioned several tens of micrometers from the flake.

The average defect intensity integrated over the 530–540 nm spectral region is plotted as a function of beam distance in Fig. S9c. The intensity shows two distinct exponential decay regimes: a rapid decay within the first 4.5 µm, followed by a slower decay extending to the large measured distances. The short-range component is consistent with local excitation close to the flake. The long-range component is attributed to substrate-assisted excitation mediated predominantly by backscattered electrons generated in the Au-coated glass substrate, which retain sufficient energy to excite the hBN defects several micrometers away from the primary electron beam[15]. These measurements show that the observed defect emission is not confined to the direct electron-beam interaction volume but contains a significant contribution from substrate-assisted excitation, which extends the effective excitation range well beyond the dimensions of the primary electron beam.

Figure S10 shows the measurements performed on hBN flakes supported on Au-TEM gird. A similar distance-dependent response is observed, with the defect emission remaining detectable as the electron beam is displaced away from the flake. The scan range is limited to shorter distances by the geometry of the sample. Local reductions in the CL intensity originate from scan positions overlapping with the Au void cavities, where little or no hBN is present. Apart from these regions, the distance dependence follows the same trend as in Fig. S9, indicating that substrate-assisted indirect excitation also contributes to the defect emission on the Au TEM grid substrate.

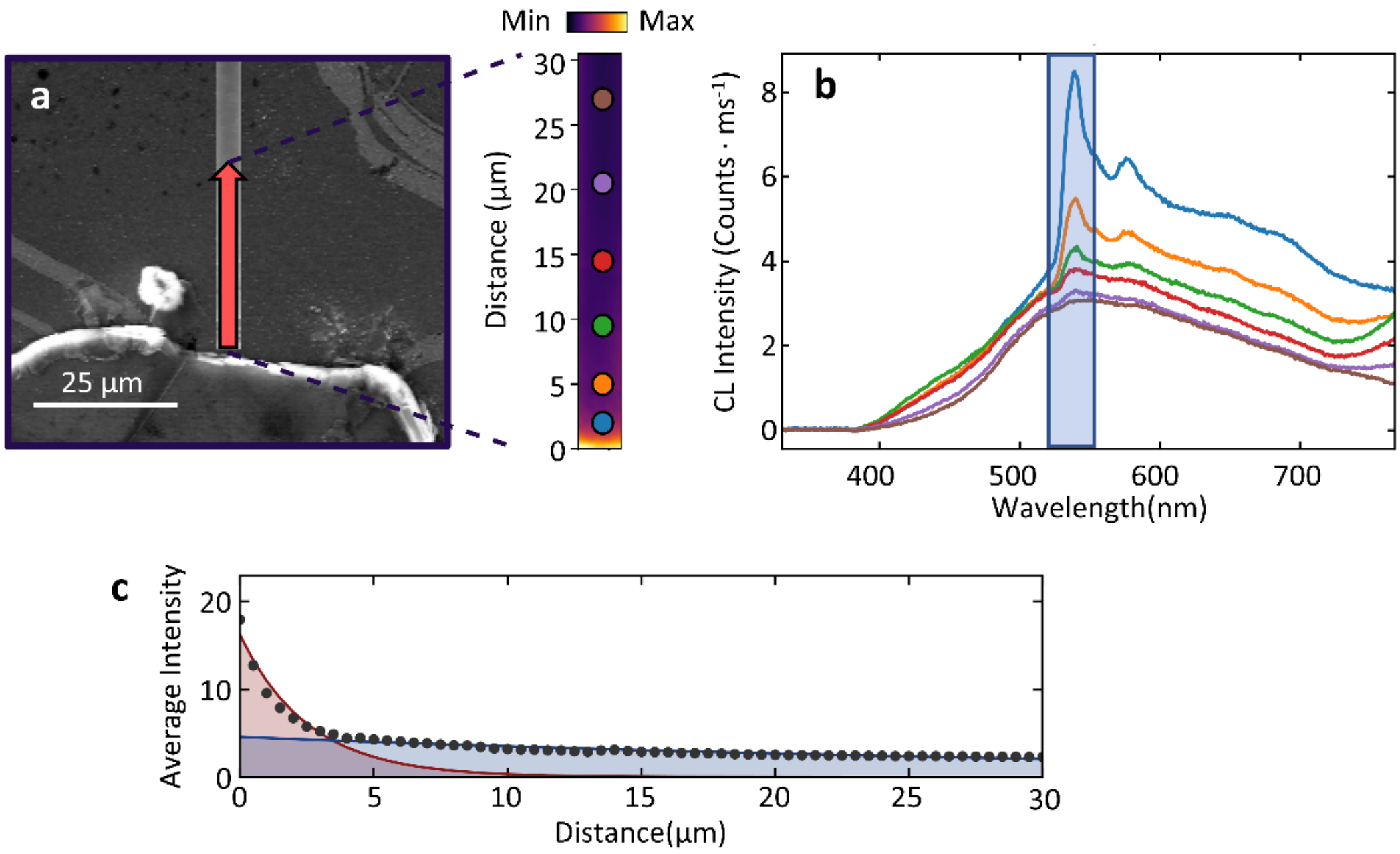

**Supplementary Fig. S12: Distance-dependent CL measurements on an hBN flake supported on Au-coated glass.** (a) SEM image showing the electron-beam scan trajectory. (b) CL spectra at different beam positions. (c) Average defect intensity (530–540 nm) as a function of distance with a two-exponential fit.

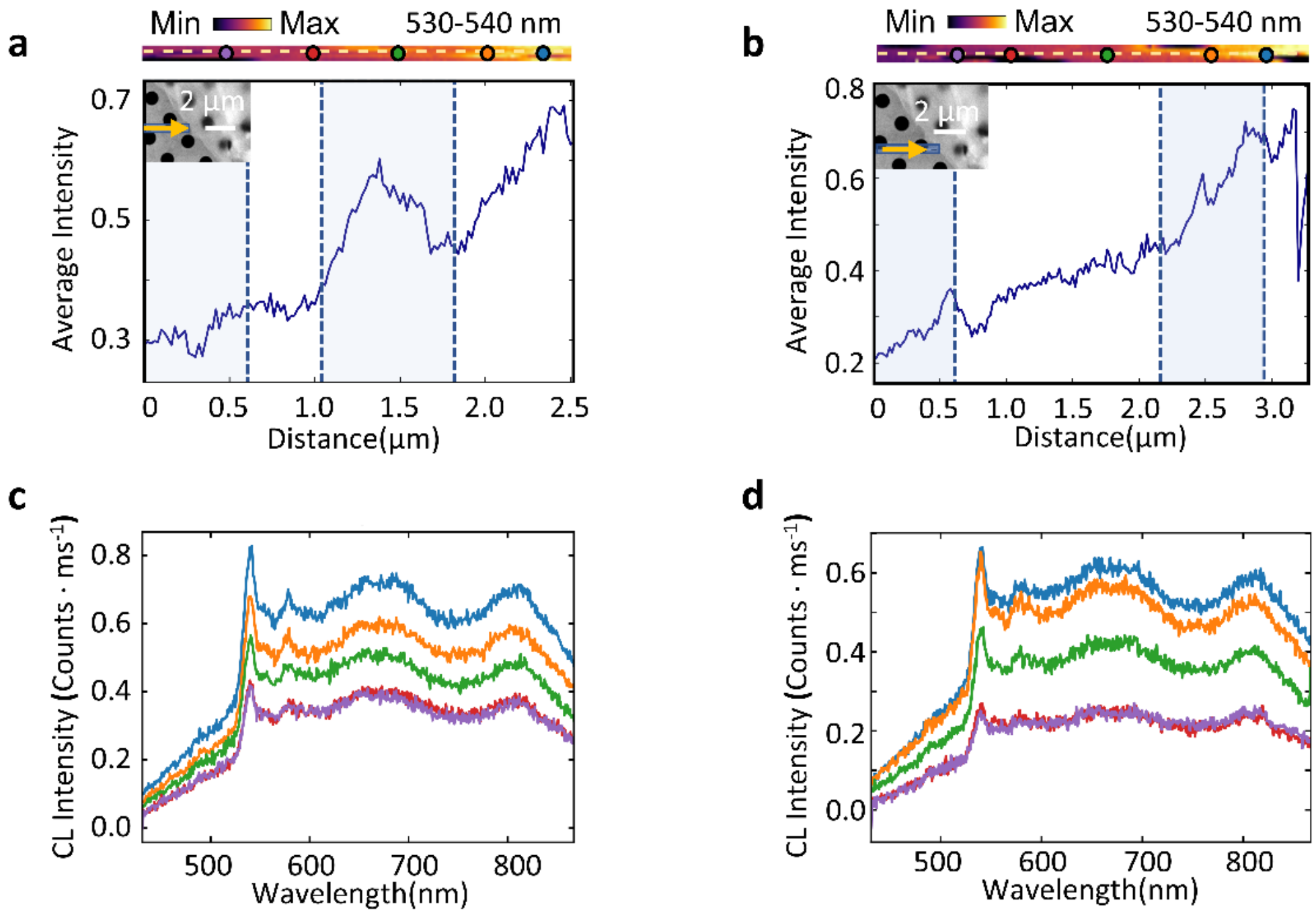

**Supplementary Figure. S13: Distance-dependent CL measurements on hBN flakes supported on an Au TEM grid.** Top panels (a,b) show the integrated defect intensity (530–540 nm) as a function of distance. Bottom panels(c,d) show the corresponding CL spectra measured at the indicated positions.

**Supplementary Note 7: Measured Reflection, Transmission, and Absorption Spectra.**

To further understand the linear response of the system and possible optical modes, we have measured the far-field reflection (R), transmission (T), and absorption (A) spectra of the pure gold grid (Supplementary Fig. S14a) and compared it with the optical responses from an hBN/Au-grid hybrid system (Supplementary Fig. S14b to d). For the latter, the responses from a thin hBN flake (~75 nm, see supplementary Fig. S15c and d) are compared to the responses from a thick hBN flake (~4 μm, see Fig. 5b), both positioned on the same gold grid.

Interestingly, regardless of the sample thickness, a dip in the A spectrum is observed, that is attributed to the hBN emitters, absorbing light at the wavelength of 537 nm, corresponding to the ZPL of the emitters. Due to the rather poor collection efficiency and sensitivity of our CCD camera at wavelengths shorter than 500nm, the signal-to-noise ratio at those wavelengths are considerably larger, therefore hindering the observation of spectral features attributed to the emitter type 1 in our system. For thicker hBN flakes, a larger number of Fabry-Pérot resonances are observed. Nevertheless, due to the low refractive index of hBN, the observed Fabry-Pérot resonances are broad and hardly resolved in the spectrum.

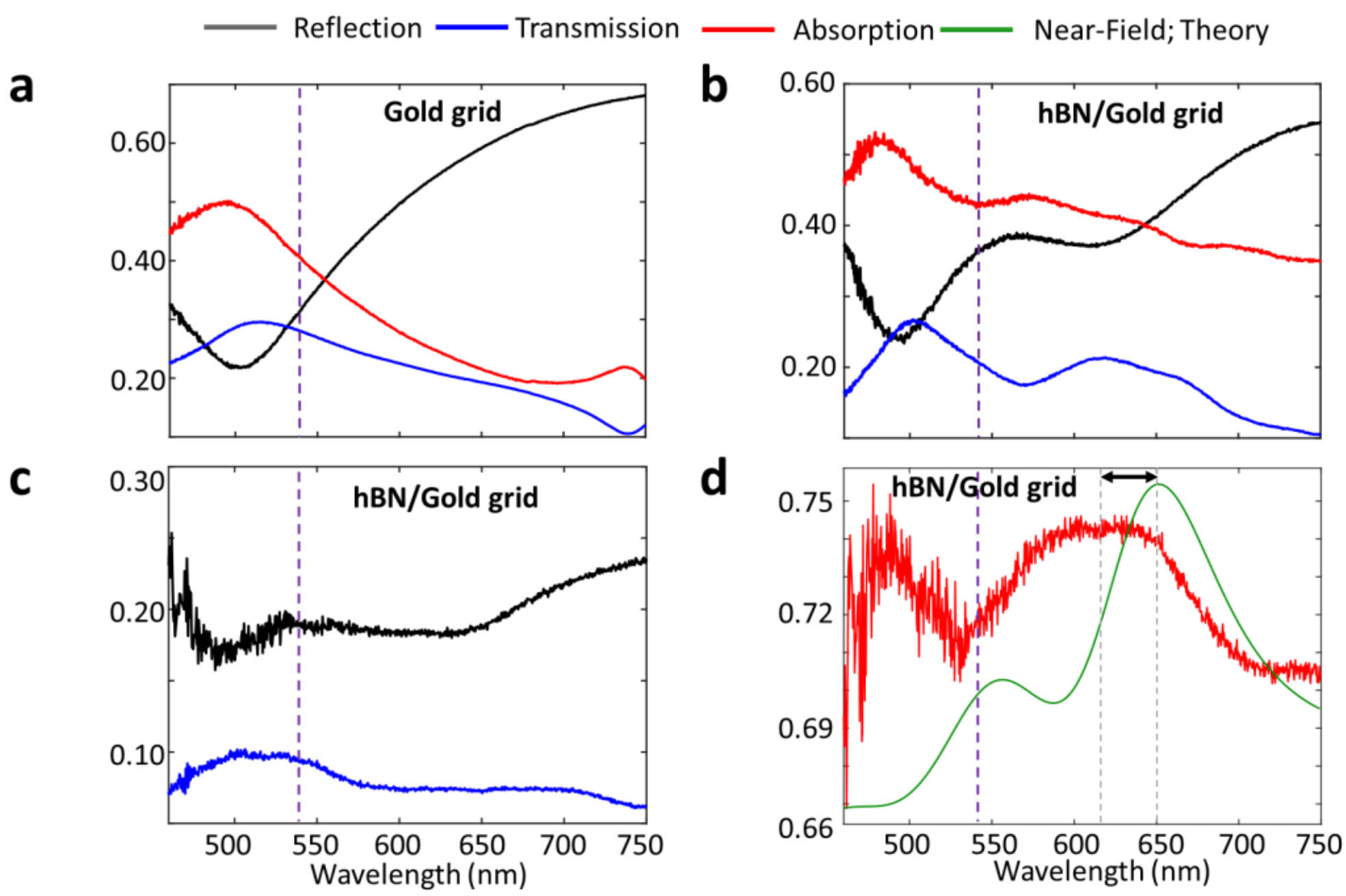

**Supplementary Figure. S14: Linear optical responses of hBN flakes positioned on the gold grid.** Measured reflection(R), transmission(T) spectra with Absorption (A) calculated as A = 1 − R − T (a) of the bare Au grid. (b) thick hBN flake on Au grid. (c) R and T of thin hBN flake on Au grid (see Supplementary Fig. S15 for the flake morphology) (d) corresponding A shown together with the near-field cavity response (green line), where the latter is calculated numerically (see Fig. 3). The dashed purple line shows the position of the ZPL associated to emitter 2.

**Supplementary Note 8: Thickness Map and Optical Image of hBN on Au Grid**

Here, we show the results of the profilometry measurements of the thin hBN flake explored in main text, positioned on the gold grid. The

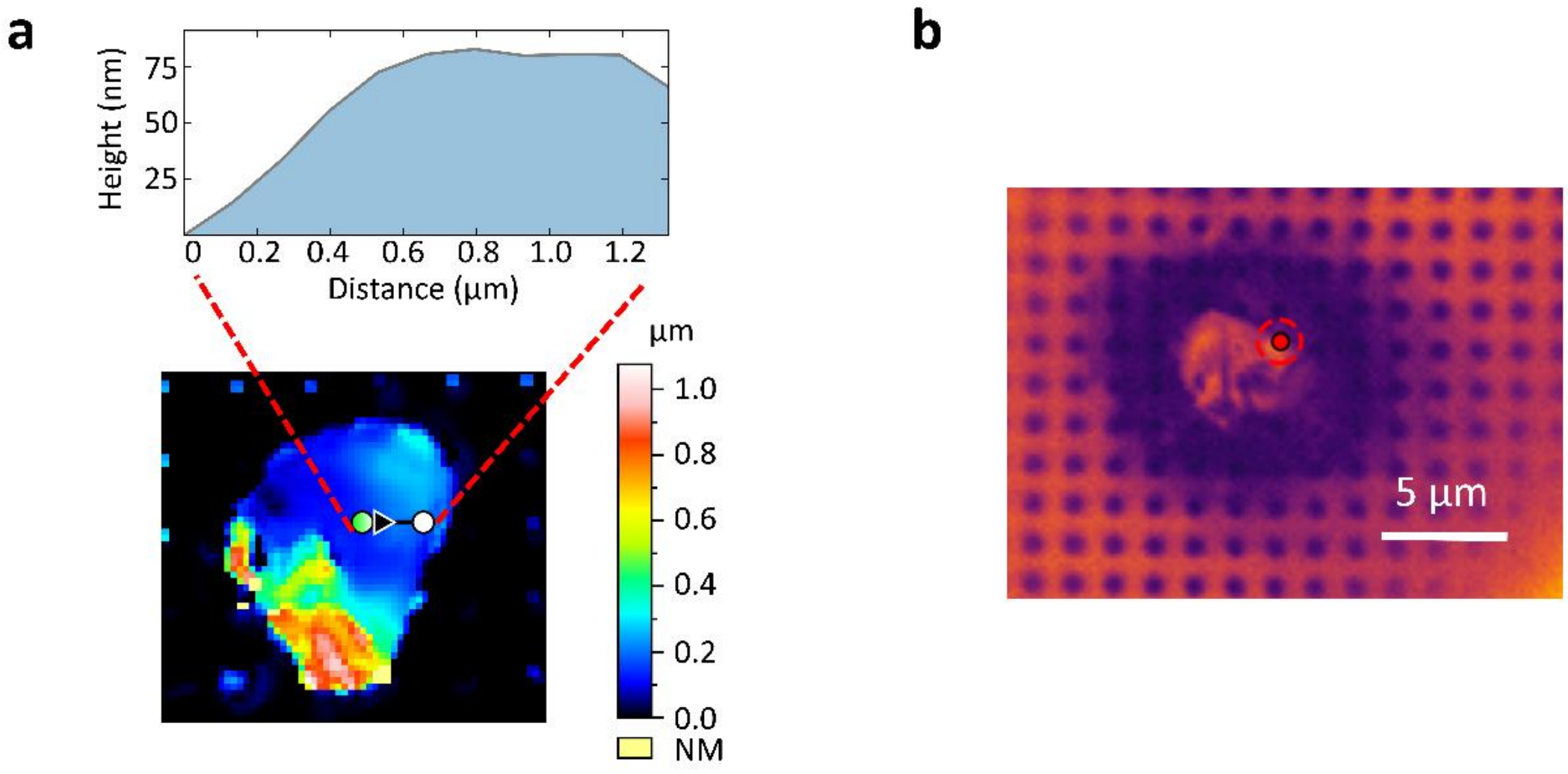

**Supplementary Figure. S15: Profilometry measurements of the flake explored in Fig. 1 to 3 of the manuscript.** (a) Thickness map and height profile. (b) Optical Microscope image of the same hBN flake on Au grid.

## References

1. Bittorf, P. H., Davoodi, F., Taleb, M. & Talebi, N. Mapping optical Bloch modes of a plasmonic square lattice in real and reciprocal spaces using cathodoluminescence spectroscopy. *Opt. Express* **29**, 34328 (2021).
2. Talebi, N. *et al. Plasmonic Grating as a Nonlinear Converter-Coupler*. *Phys. Rev. Lett* vol. 93 www.eecs.wsu.edu/~schneidj/ufdtd. (2004).
3. Talebi, N. *et al.* Merging transformation optics with electron-driven photon sources. *Nature Communications* **10**, (2019).
4. Ögüt, B., Talebi, N., Vogelgesang, R., Sigle, W. & Van Aken, P. A. Toroidal plasmonic eigenmodes in oligomer nanocavities for the visible. *Nano Lett.* **12**, 5239–5244 (2012).
5. Talebi, N. Interaction of electron beams with optical nanostructures and metamaterials: from coherent photon sources towards shaping the wave function. *Journal of Optics* **19**, 103001 (2017).
6. Talebi, N., Mahjoubfar, A. & Shahabadi, M. Plasmonic ring resonator. *Journal of the Optical Society of America B* **25**, 2116 (2008).
7. Talebi, N., Sigle, W., Vogelgesang, R. & Van Aken, P. Numerical simulations of interference effects in photon-assisted electron energy-loss spectroscopy. *New J. Phys.* **15**, (2013).
8. Etchegoin, P. G., Le Ru, E. C. & Meyer, M. An analytic model for the optical properties of gold. *Journal of Chemical Physics* **125**, (2006).
9. Etchegoin, P. G., Le Ru, E. C. & Meyer, M. Erratum: "An analytic model for the optical properties of gold" [J. Chem. Phys. 125, 164705 (2006)]. *J. Chem. Phys.* **127**, (2007).
10. Talebi, N. *et al.* Excitation of Mesoscopic Plasmonic Tapers by Relativistic Electrons: Phase Matching versus Eigenmode Resonances. *ACS Nano* **9**, 7641–7648 (2015).

11. Talebi, N. A directional, ultrafast and integrated few-photon source utilizing the interaction of electron beams and plasmonic nanoantennas. *New J. Phys.* **16**, (2014).

12. Ding, S. Y. *et al.* Nanostructure-based plasmon-enhanced Raman spectroscopy for surface analysis of materials. *Nature Reviews Materials* vol. 1 Preprint at https://doi.org/10.1038/natrevmats.2016.21 (2016).

13. Singh, P., Garg, B., Black, M., Taleb, M. & Talebi, N. Defect-mediated inelastic light scattering by infrared phonons in hexagonal boron nitride. *under revision* (2026).

14. Islam, M. S. *et al.* Large-Scale Statistical Analysis of Defect Emission in hBN: Revealing Spectral Families and Influence of Flake Morphology. *ACS Nano* **18**, 20980–20989 (2024).

15. Ebel, S., Mortensen, N. A. & Morozov, S. Substrate-assisted cathodoluminescence. *NPJ Nanophotonics* **3**, (2026).